\documentclass[aps,pra,10pt,onecolumn,superscriptaddress,nofootinbib]{revtex4-2}
\usepackage{graphicx} 
\usepackage{amsfonts}
\usepackage{dsfont}
\usepackage{mathtools}
\usepackage{amssymb}
\usepackage{csquotes}
\usepackage{xcolor}
\usepackage{orcidlink}
\usepackage{tikz-cd}
\usepackage{enumitem}
\usepackage{hyperref}
\usepackage{natbib}
\newcommand{\tra}[1]{\text{tr}\left(#1\right)}
\newcommand{\ri}{\rightarrow}
\newcommand{\R}{\mathbb{R}}

\newcommand{\af}{\alpha}
\newcommand{\ketbra}[1]{\left| #1 \rangle \langle #1 \right| }

\newcommand{\Exp}[1]{\left\langle #1 \right\rangle_{\psi_{t}}}

\newtheorem{definition}{Definition}
\newtheorem{lemma}{Lemma}
\newtheorem{proposition}{Proposition}
\newtheorem{theorem}{Theorem}
\newtheorem{corollary}{Corollary}

\usepackage[capitalise,nameinlink]{cleveref}
\usepackage[authormarkuptext=name,commentmarkup=uwave]{changes}
\definechangesauthor[name={AT}, color=red]{alisson}
\definechangesauthor[name={JRH},color=green!70!black]{jonte}
\definechangesauthor[name={NEW}, color=blue!70]{rev}

\begin{document}

\title{Sufficiency of the counterfactual account of L\"uders' rule to rule out ontological models of quantum mechanics}

\author{Alisson Tezzin\,\orcidlink{0000-0002-5849-0124}}
\email{alisson.cordeiro@ime.eb.br}
\affiliation{Military Institute of Engineering (IME), Defense Engineering Section, Rio de Janeiro, Brazil}
\affiliation{QuIIN — Quantum Industrial Innovation, EMBRAPII CIMATEC Competence Center in Quantum Technologies, SENAI CIMATEC, Av. Orlando Gomes, 1845, Salvador, Bahia, 41850-010, Brazil}
\affiliation{Department of Mathematical Physics, Institute of Physics, University of São Paulo, Rua do Matão 1371, São Paulo 05508-090, São Paulo, Brazil}
\author{B\'arbara Amaral\,\orcidlink{0000-0003-1187-3643}}
\affiliation{Department of Mathematical Physics, Institute of Physics, University of São Paulo, Rua do Matão 1371, São Paulo 05508-090, São Paulo, Brazil}
\author{Jonte R. Hance\,\orcidlink{0000-0001-8587-7618}}
\email{jonte.hance@newcastle.ac.uk}
\affiliation{School of Computing, Newcastle University, 1 Science Square, Newcastle upon Tyne, NE4 5TG, UK}
\affiliation{Quantum Engineering Technology Laboratories, Department of Electrical and Electronic Engineering, University of Bristol, Woodland Road, Bristol, BS8 1US, UK}

\date{\today}

\begin{abstract}
Ontological models, as used in the generalised contextuality literature, play a central role in current research on quantum foundations, providing a framework for defining classicality, constructing classical analogues of key quantum phenomena, and examining the ontology of quantum states. In this work, we show that a counterfactual account of L\"uders’ rule—-- which we argue is naturally implied by the mathematical structure of the rule itself —--renders such models inherently incompatible with the quantum formalism. This incompatibility arises because the counterfactual update requires ontological models to update their states according to conditional probability, which in turn which in turn renders predictions of sequential measurements order-independent. This implies that ontological models, even contextual ones, must either act differently to what we would expect given (this, typically implicitly-assumed account of) quantum state update rule, or cannot model quantum behaviour.
\end{abstract}
\maketitle

\tableofcontents

\section{Introduction}

One of the most debated fundamental problems in modern physics is whether quantum formalism's description of nature is complete, or if it could be supplemented with additional, or ``hidden'', variables \cite{einstein1935complete, bohr1935complete, bell1966problem, kochen1975problem, bohm1952hiddenI, broglie1960wave, belinfante1973survey}. While hidden variable theories have been constructed which (partially or fully) reproduce the predictions of quantum mechanics \cite{bohm1952hiddenI, bohm1852hiddenII, wiener1955differentialTheory, belinfante1973survey}, various impossibility theorems, notably Bell's \cite{bell1964epr, bell1985beables} and Kochen and Specker's \cite{kochen1975problem}, have shown that such attempts inevitably violate certain consistency conditions which we would arguably want from a successful description of reality \cite{neumann1934algebraic, jauch1963hidden, bell1964epr, kochen1975problem}. Bell demonstrated the impossibility of reproducing the predictions of quantum mechanics with those \textit{local} hidden variable theories which also obey statistical independence (i.e., where our choice of which measurement(s) to perform in each run of an experiment is in no way correlated with said hidden variable \cite{hance2022localBell,Hance2024CounterfactualRestriction}). Kochen and Specker, on the other hand, proved the impossibility of assigning values to all quantum observables in a way that preserves functional relations between them (e.g., if the value of the physical quantity $\hat{A}$ is $\af$, where the value of $\hat{A}^{2}$ is $\af^{2}$, etc) \cite{kochen1975problem, isham1998toposI, landsman2017foundations}. This obstructs what we now call \textit{noncontextual} hidden variable descriptions \cite{abramsky2011sheaf, landsman2017foundations, budroni2022contextualityReview}. Bell's and Kochen and Specker's seminal works gave rise to two major modern research areas: nonlocality \cite{brunner2014bell, popescu2014beyond, Popescu1994axiom} and contextuality \cite{budroni2022contextualityReview, amaral2018graph, spekkens2005contextuality}.

Research in quantum foundations, and on  contextuality and nonlocality in particular, has recently been dominated by an operational approach to physics \cite{fuchs2000noInterpretation,plavala2023general, fuchs2010qbism, markus2021probabilistic, popescu2014beyond, budroni2022contextualityReview, spekkens2005contextuality, abramsky2011sheaf}. This approach treats operational notions (such as preparations, transformations, and measurements) as the basic concepts of physical theories, and views the role of such theories as to specify probabilities of events associated with these operations \cite{spekkens2005contextuality, popescu2014beyond, plavala2023general}. To accommodate contextuality in this framework, Spekkens reinterpreted it as the impossibility of assigning the same ontological representation for operationally equivalent experimental procedures, resulting in his notion of generalised contextuality \cite{spekkens2005contextuality, spekkens2019Leibniz}. Ontological representations of experimental procedures are defined within what are known as ontological models, a form of hidden variable model which it is claimed generalises Kochen and Specker's formulation to cover any operational theory \cite{schmid2021characterization}. The framework in which these models are defined is referred to as the Ontological Models Framework (OMF). The ultimate aim of this generalised contextuality research programme is to establish a universally applicable notion of classicality \cite{schmid2021characterization, schmid2024structureTheorem, schmid2024inequalities, schmid2024objections}. Generalised contextuality quickly expanded as a research field following Spekkens' original paper, sparking extensive investigations into geometric analyses, noncontextuality inequalities, polytopes, and other analytical tools to study and understand its principles \cite{selby2023witnessing,Wagner2023engineerContextuality,schmid2024structureTheorem,schmid2024inequalities}. It has also been linked to significant phenomena in thermodynamics \cite{comar2024anomolous, lostaglio2020quantumSignatures}, quantum Darwinism \cite{baldijao2021darwinism}, cloning \cite{lostaglio2020cloning}, quantum interrogation \cite{Wagner2024coherence}, and other areas.

In this paper, we explore the quantum state update rule (L\"uders' rule \cite{luders1950upate}) to establish and equivalence between the pairwise compatibility of all observables in a scenario, and the existence of an ontological model for that scenario. This extends our recent result \cite{Tezzin2025deterministicModels} for deterministic models to stochastic ones. We focus on ontological models because, similarly to that case, measurements are thought of updating ``one’s information about what the ontic state of the system was prior to the procedure being implemented'' \cite{spekkens2005contextuality}, whereas quantum states are treated as statistical mixtures of (i.e., probability measures over) such underlying states. Additionally, quantum states (and wavefunctions in particular) are viewed as an ``ensemble average'' of real ontic states of the system, which behave in ways we would expect real states of the world to --- unlike wider hidden variable models, which can nominally be as contrived as we like \cite{norsen2017foundations,caticha2021hamiltonKillingFlows,caticha2025entropicDynamicsApproach,waegell2023localQuantumTheory,Waegell2024madelungSpacetime}. Since we consider Lüders' rule, our analysis does not apply --- at least not straightforwardly --- to hidden-variable reformulations of quantum mechanics that directly \cite{Hance2022MeasProb} exclude the collapse postulate, such as the de Broglie-Bohm pilot-wave theory \cite{broglie1960wave,bohm1952hiddenI,bohm1852hiddenII,norsen2017foundations}. 

As in Ref.~\cite{Tezzin2025deterministicModels}, we start by defending a counterfactual account of L\"uders' rule:  instead of being updated by propositions in the simple past (``a measurement of $\hat{A}$ \textit{was} made, and the value $\af$ \textit{was} obtained''), we argue that quantum states are theoretically updated to ensure the validity of propositions in the first conditional (``\textit{if} a measurement is made of $\hat{A}$, the result \textit{will be} found to be $\af$'') \cite{doring2011what}. This is to say that, contrary to the typical description of updated states as \textit{post-measurement states},  updated states  are theoretically constructed as \textit{pre}-measurement states, with the qualifier \textit{post}-measurement state regarded instead as a merely circumstantial attribute. This approach aligns with multiple perspectives on quantum mechanics and, as we show, is supported by the mathematical formulation of L\"uders' rule itself \cite{nakamura1962neumann, bub1977conditionalization, earman2019conditionalization, mermin2022noProblem,mueller2024idealism}. We then argue that. by this account, L\"uders' rule translates to ontological models as conditional probability, because considerations about interactions with measuring apparatuses are automatically removed in this counterfactual description. Using Bayes' rule and Kolmogorov's extension theorem \cite{tao2011measure}, we show that a set of quantum observables admits a state-updating ontological model (where states are updated via conditional probability) if and only if that set consists of pairwise compatible observables. In particular, a single pair of incompatible observables is sufficient to rule out a scenario being representable by an ontological model. The impossibility of modelling incompatible observables with state-updating ontological models follows from the order-dependence of quantum predictions for sequential measurements of incompatible observables, in contrast with predictions given by state-updating ontological models. 

\citet{fine1982bell,fine1982commutingObservables} and \citet{malley2004commuteSimultaneously}, separately and together \cite{malley2005localRealism} have previously argued that incompatibility is sufficient to rule out deterministic underlying states. \citet{fine1982bell,fine1982commutingObservables} demonstrated that underlying-state models that correctly predict joint distributions of observables, along with the marginals of their functions, exist only for compatible observables (Theorem 7 of Ref.~\cite{fine1982commutingObservables}); he also famously showed that underlying-state models exist if and only if there are well-defined joint distributions for pairs of non-commuting observables in the standard Bell scenario \cite{fine1982bell}. \citet{malley2004commuteSimultaneously}, in turn, proved that in Hilbert spaces of dimension greater than 2, underlying-state models that correctly predict joint distributions of pairs of compatible observables (a form of noncontextuality \cite{amaral2018graph,abramsky2011sheaf}) must update states via conditional probability, which in turn obstructs incompatible observables. Here, we essentially take the opposite route: we argue that the mathematical formulation of the quantum state update rule entails that ontological models must obey conditional probability, from which it follows that incompatibility obstructs their existence. As shown in Lemma \ref{lemma:ontologicalModel} and Corollary \ref{cor:DeterministicModel}, Fine and Malley's consistency conditions follows from ours.

\citet{harrigan2007interpretation} indirectly explores the realisation that ontological (or hidden variable) models cannot accurately update states via conditional probability. They suggest that this limitation is an expected property, justified by the interactions between the measured system and the measuring apparatus. Similar reasoning appears in other works where ontological models deliberately fail to respect conditional probability \cite{Hindlycke2022ConOntologicalModel}, such as the so-called epistemically restricted theories \cite{spekkens2007toy,vanEnk2007,Paterek2010Limited,Coecke2011Toy,Bartlett2012EpistemticRestricted,Spekkens2016Epirestricted,Budiyono2019Epirestrict,Braasch2022Epirestricted,Surace2024InaccessibleInfo,Fankhauser2025EpistHorizon}. This perspective, however, conflicts with our counterfactual description of the state update. We therefore take the opposite stance, arguing that such an impossibility demonstrates the inadequacy of ontological models for describing the ontology of quantum systems. This is the same conclusion Spekkens draws from quantum generalised contextuality \cite{spekkens2019Leibniz}, differing only in the consistency condition that causes the inadequacy.

\citet{Selby2023ContextwoutIncompat} showed in a recent paper that one can violate generalised noncontextuality without requiring incompatibility, by making use of the fact that the elements of a POVM, despite potentially being non-orthogonal, technically count as compatible. The introduction of Ref.~\cite{Selby2023ContextwoutIncompat} further claims that measurement incompatibility is neither necessary nor sufficient for a generalised-noncontextual ontological model to be impossible. Our result shows this is not the case (assuming you accept our argument that ontological models should state-update through conditional probability), countering the ``not sufficient'' part of this claim. Ref.~\cite{Hance2025Fluctuations} focuses on issues with the meaningfulness of the ``not necessary'' part of this claim.

This paper is laid out as follows. First, in Section~\ref{sec:notation}, we go over the notation we use. In Section~\ref{sec:stateUpdate}, we defend a counterfactual account of L\"uders' rule, showing that its very the mathematical formulation suggests that, at the theoretical level, quantum states are updated by counterfactual propositions. In Section~\ref{sec:compatibility}, we look at compatibility of observables, before showing that there is a latent tension between how the quantum formalism models state update after measuring incompatible observables, and how state update is often rendered in hidden-variable models. In Section~\ref{sec:omf}, we introduce the Ontological Models Framework, first in its standard form, then in the form it needs to take to reproduce the quantum state update rule in our counterfactual account. In Section~\ref{sec:results} we give the main result of our paper---that, under our counterfactual account, for any ontological model of the form needed to reproduce L\"uders' rule, all observables measurable in that model must be pairwise compatible. In Section~\ref{sec:disc} we then discuss the assumptions behind our result, and the consequences of it.

\section{Notation}\label{sec:notation}

We first recall the basic formalism of quantum theory, and establish the notation that will be used throughout this paper.  Precise definitions can be found in Refs.~\cite{hannabuss1997introduction, hall2013quantum,landsman2017foundations} and elsewhere. The measure-theoretic concepts we use to discuss ontological models are summarised in Appendix~\ref{sec:measureTheory}. They can also be found in standard textbooks on measure and probability theories \cite{tao2011measure, halmos1974measure,klenke2014probability,cohn2013measure}.

As is typical in modern quantum foundations, we will give our results for finite-dimensional systems only. However, we would expect these results naturally extend to infinite-dimensional quantum systems. Future work will involve proving this formally.

As usual, we will follow the Hilbert space formalism of quantum theory \cite{hannabuss1997introduction,hall2013quantum}. In this approach, associated with a finite-dimensional quantum system $\mathfrak{S}$ is a finite-dimensional complex Hilbert space $\mathcal{H}$ \cite{hall2013quantum}. The \textit{physical quantities} (also called \textit{observables}) of $\mathfrak{S}$ are represented by self-adjoint operators in $\mathcal{H}$. \textit{States} of $\mathfrak{S}$ are density operators in $\mathcal{H}$, namely positive self-adjoint operators with trace $1$. The sets of observables and states of $\mathfrak{S}$ will be denoted by $\mathcal{O} \equiv \mathcal{B}(\mathcal{H})_{\text{sa}}$ and $\mathcal{S}$, respectively.

$\Exp{ \ \cdot \ }$ denotes the expectation defined by a state $\hat{\rho}$. That is, $\Exp{ \ \cdot \ }$ is the linear functional on the set $\mathcal{B}(\mathcal{H})$ of operators given by
\begin{equation}
    \Exp{\hat{A}} \doteq\label{eq:quantumExpectationFinite} \tra{\hat{\rho} \hat{A}}
\end{equation}
for each $\hat{A} \in\mathcal{B}(\mathcal{H})$. If $\hat{A}$ is an observable, $\Exp{\hat{A}}$ is the expected value of $\hat{A}$ for the state $\hat{\rho}$.

Given a state $\hat{\rho}$ and an observable $\hat{A}$, we denote by $P_{\hat{\rho}}[\hat{A}=  \cdot \ ]$ the distribution of $\hat{A}$ in the state $\hat{\rho}$. That is, $P_{\hat{\rho}}[\hat{A}=  \cdot \ ]$ is the probability distribution on the spectrum $\sigma(\hat{A})$ of $\hat{A}$ given by
\begin{equation}
    P_{\hat{\rho}}[\hat{A}=\af] \doteq\label{eq:quantumProbabilityValue} \Exp{\hat{\Pi}(\hat{A}=\alpha)},
\end{equation}
where $\hat{\Pi}(\hat{A}=\alpha)$ is the projection onto the subspace of $\mathcal{H}$ spanned by the eigenvalue $\af$ of $\hat{A}$. Similarly, $P_{\hat{\rho}}[\hat{A} \in  \cdot \ ]$ denotes the probability measure on (the power set of) $\sigma(\hat{A})$ induced by $P_{\hat{\rho}}[\hat{A}=  \cdot \ ]$. This means, for any set $\Delta \subset \sigma(\hat{A})$,
\begin{equation} \begin{split}
    P_{\hat{\rho}}[\hat{A} \in \Delta] &\doteq\label{eq:quantumProbabilitySet} \sum_{\af \in \Delta} P_{\hat{\rho}}[\hat{A}=\af]
    \\
    &= \Exp{\hat{\Pi}(\hat{A}\in\Delta)},
\end{split}\end{equation}
with $\hat{\Pi}(\hat{A}\in\Delta) \doteq \sum_{\af \in \Delta}\hat{\Pi}(\hat{A}=\alpha)$. Physically, $P_{\hat{\rho}}[\hat{A} \in \Delta]$ represents the probability that a measurement of $\hat{A}$ in the state $\hat{\rho}$ yields a value lying in $\Delta$.

The set $\mathcal{O}$ of observables is closed under functional relations. It means that, for any observable $\hat{A} \in \mathcal{O}$ and any function $g:\sigma(\hat{A}) \ri \R$, there exists an observable representing the physical quantity $g(\hat{A})$ (e.g., if $g$ is the function $\af \mapsto \af^{2}$, $g(\hat{A})$ is the observable $\hat{A}^{2}$). $g(\hat{A})$ is defined by
\begin{equation}
    g(\hat{A}) \doteq \sum_{\af \in \sigma(\hat{A})} g(\af)\hat{\Pi}(\hat{A} = \alpha),
\end{equation}
where $\hat{A}=\sum_{\af \in \sigma(\hat{A})}\af \hat{\Pi}(\hat{A} = \alpha)$ is the spectral decomposition of $\hat{A}$. Recall that, in particular, $\hat{\Pi}(\hat{A} \in \Delta) = \chi_{\Delta}(A)$, where $\chi_{\Delta}:\sigma(\hat{A}) \ri \{0,1\}$ is the indicator function of $\Delta$.

The spectrum of $g(\hat{A})$ is the image of $\sigma(\hat{A})$ under $g$, i.e., $\sigma(g(\hat{A}))=g(\sigma(\hat{A}))$ \cite{kadison1983fundamentals}. The distribution of $g(\hat{A})$ in some state $\hat{\rho}$ is the distribution of the random variable $g$ w.r.t. $P_{\hat{\rho}}[\hat{A} \in   \cdot \ ]$. It means that, for each $\Sigma \subset \sigma(g(\hat{A}))$,
\begin{equation}
    P_{\hat{\rho}}[g(\hat{A}) \in \Sigma] =\label{eq:pushforward} P_{\hat{\rho}}[\hat{A} \in g^{-1}(\Sigma)].
\end{equation}
Finally, the expectation of $g(\hat{A})$ is the expected value of $g$ w.r.t. $P_{\hat{\rho}}[\hat{A} \in  \cdot \ ]$:
\begin{equation}
    \Exp{g(\hat{A})} =\label{eq:functionExpectation} \sum_{\af \in \sigma(\hat{A})} g(\af) P_{\hat{\rho}}[\hat{A}=\af].
\end{equation}

The set of states $\mathcal{S}$ is a convex subset of $\mathcal{B}(\mathcal{H})$. It means that, if $\hat{\rho}_{1},\hat{\rho}_{2}$ are states and $p \in [0,1]$, the convex combination
\begin{equation}
    p\hat{\rho}_{1} + (1-p)\hat{\rho}_{2}
\end{equation}
is also a state. A state is said to be \textit{pure} if it is an extreme point of $\mathcal{S}$, meaning it cannot be written as a convex combination of two distinct states. $\mathcal{S}$ is the convex hull of the set of pure states, which means that every state is a convex combination of pure states.

In finite-dimensional systems, pure states correspond to rank-1 projections or, equivalently, to one-dimensional subspaces. For historical and conceptual reasons, pure states are often identified with the unit vectors that generate the subspaces they correspond to. That is, if $\psi \in \mathcal{H}$ is a unit vector, we say that $\psi$ is a pure state, in allusion to the subspace this vector spans. We denote by $\mathcal{S}_{0}$ the set of all pure states in $\mathfrak{S}$.

As usual, $\langle \cdot \vert \cdot \rangle $ denotes the inner product of a Hilbert space $\mathcal{H}$, and $\Vert \ \cdot \ \Vert$ denotes the corresponding norm. That is, for any $\psi \in \mathcal{H}$ we have
\begin{equation}
    \Vert \psi\Vert \doteq \sqrt{\langle \psi | \psi \rangle}.
\end{equation}

In Dirac notation, the projection onto the subspace spanned by a unit vector $\psi$ is denoted by $\ketbra{\psi}$. For simplicity, we will denote by $P_{\psi}[\hat{A} = \cdot \ ]$ the distribution of the observable $\hat{A}$ in the state $\ketbra{\psi}$ (analogously for the associated probability measure). For any unit vector $\psi$ and any observable $\hat{A}$,
\begin{equation} \begin{split}
    \langle\hat{A}\rangle_{\psi} &= \tra{\ketbra{\psi} \hat{A}}
    \\
    &= \langle \psi | \hat{A} | \psi \rangle.
\end{split}\end{equation}
In particular, for any pure state $\psi$, observable $\hat{A}$ and set $\Delta \subset \sigma(\hat{A})$,
\begin{equation} \begin{split}
    P_{\psi}[\hat{A} \in \Delta] &= \tra{\ketbra{\psi} \hat{\Pi}(\hat{A} \in \Delta)}
    \\
    &= \langle \psi | \hat{\Pi}(\hat{A} \in \Delta) | \psi \rangle.
\end{split}\end{equation}

Another important concept in quantum information is that of positive operator-valued measures (POVMs) \cite{nielsen2000quantum}. These objects represent measurements that do not necessarily have a direct and idealised correspondence with physical quantities (a POVM in $\mathcal{H}$ can be any arbitrary finite set $\{\hat{E}_{i}\}_{i \in I}$ of positive self-adjoint operators that sum to the identity). Since one goal of generalised contextuality is to provide a framework that accommodates such generalised measurements, the literature often relies on POVMs rather than observables. Here though, we will focus exclusively on observables, as this is sufficient for proving our main result (and our results should extend naturally to sequential measurements of ``incompatible'' POVMs).

\section{State Update in the Quantum Formalism}\label{sec:stateUpdate}

Let $\hat{A}$ be an observable and $\Delta$ be a set of its eigenvalues. Consider the counterfactual proposition ``If a measurement is made of $\hat{A}$, the result will be found to lie in $\Delta$'' \cite{doring2011what}, denoted $[\hat{A} \in \Delta]$ hereafter. These are the propositions regarding observables that physical theories (at least at an instrumentalist level) are designed to handle, as they correspond to assertions about future (i.e., subsequent) measurements results.

In quantum mechanics, states assign probabilities to propositions  through the Born rule, which involves taking the trace of the product of the state $\hat{\rho}$ with the projection $\hat{\Pi}(\hat{A}=\Delta)$ (Eqs.~\eqref{eq:quantumExpectationFinite} and \eqref{eq:quantumProbabilityValue}). Likewise, the quantum state update (L\"uders' rule) provides an optimal reconstruction of the state $\hat{\rho}$ ensuring the validity of the proposition $[\hat{A} \in \Delta]$ (i.e., ensuring that a value $\alpha$ lying in $\Delta$ will be obtained if $\hat{A}$ is measured). This section examines the details of the state update mechanism to clarify this point. This analysis presents an account of the theoretical quantum update mechanism. It should not be misunderstood as a particular interpretation of the state collapse that occurs when quantum systems are measured in actual experiments. 

Let us begin with the case where $\Delta$ contains only one eigenvalue $\af$. That is, $[\hat{A} = \af] \equiv [\hat{A} \in \{\af\}]$ is the proposition ``If a measurement is made of $\hat{A}$, the result will be found to be $\af$''. Let $\hat{\Pi}(\hat{A}=\alpha)$ be the corresponding projection, and let $\hat{\Pi}(\hat{A}=\alpha)(\mathcal{H})$ be the subspace of $\mathcal{H}$ onto which $\hat{\Pi}(\hat{A}=\alpha)$ projects, i.e., $\hat{\Pi}(\hat{A}=\alpha)(\mathcal{H}) = \{\hat{\Pi}(\hat{A}=\alpha)\psi: \psi \in \mathcal{H}\}$.

As expressed in Eq.~\eqref{eq:quantumProbabilityValue}, the expectation value of the projection $\hat{\Pi}(\hat{A}=\alpha)$  gives the probability that the proposition $[\hat{A} = \af]$ is true (with the collection of all propositions $[\hat{A}=\af'],\af' \in \sigma(\hat{A}),$ forming the underlying sample space). Pure states (thought of as unit vectors) that lie within the subspace $\hat{\Pi}(\hat{A}=\alpha)(\mathcal{H})$ ensure the proposition is true, in that they assign probability $1$ to $[\hat{A}=\af]$. This probability is strictly smaller than $1$ for any other pure state. More broadly, a state $\hat{\rho}$ assigns probability $1$ to $[\hat{A}=\af]$ if and only if it is a convex combination of pure states lying in the corresponding subspace.

The orthocomplement of $\hat{\Pi}(\hat{A} = \alpha)(\mathcal{H})$, namely the set
\begin{equation}
    \hat{\Pi}(\hat{A} = \alpha)(\mathcal{H})^{\perp} \doteq\label{eq:orthocomplement} \{\phi \in \mathcal{H}: \forall_{\psi \in \hat{\Pi}(\hat{A} = \alpha)(\mathcal{H})}\langle \phi\vert \psi \rangle = 0\},
\end{equation}
is the subspace corresponding the proposition $\neg[\hat{A} \in \af] \equiv [\hat{A} \neq \af] \equiv [\hat{A} \in \sigma(\hat{A})\backslash \{\af\}]$, which asserts that ``If a measurement is made of $\hat{A}$, the result will be found to be other than $\af$''. This is because a pure state $\psi$ satisfies $P_{\psi}[\hat{A} \in \sigma(\hat{A})\backslash\{\af\}]=1$ if and only if $\psi \in \hat{\Pi}(\hat{A} = \alpha)(\mathcal{S})^{\perp}$. These two subspaces have trivial intersection (i.e., their intersection is the zero vector) and their direct sum is the entire Hilbert space \cite{kadison1983fundamentals}. It means that, for any vector $\psi\in \mathcal{H}$, there exists a unique pair $\psi_{\af} \in \hat{\Pi}(\hat{A} = \alpha)(\mathcal{H})$, $\psi_{\neg\af} \in \hat{\Pi}(\hat{A} = \alpha)(\mathcal{H})^{\perp}$, such that
\begin{equation}
    \psi = \psi_{\af} + \psi_{\neg\af}.
\end{equation}
Furthermore, $\psi_{\af}$ is the unique element of $\hat{\Pi}(\hat{A} = \alpha)(\mathcal{H})$ that minimises the distance from $\psi$, in that
\begin{equation}
    \Vert \psi - \psi_{\af}\Vert =\label{eq:minimizeDistance} \min \{\Vert \phi - \psi\Vert: \phi \in \hat{\Pi}(\hat{A} = \alpha)(\mathcal{H})\},
\end{equation}
and analogously for $\psi_{\neg\af}$. The vector $\psi_{\af}$ is said to be the orthogonal projection of $\psi$ on $\hat{\Pi}(\hat{A} = \alpha)(\mathcal{H})$. It follows by construction that
\begin{equation}
    \psi_{\af} = \hat{\Pi}(\hat{A} = \alpha) \psi.
\end{equation}

Projecting a pure state $\psi$ onto $\hat{\Pi}(\hat{A} = \alpha)(\mathcal{H})$ is an optimal way of reconstructing $\psi$ to ensure the validity of the proposition $[\hat{A} =\af]$, i.e., to ensure --- at the theoretical level --- that, if a measurement is made of $\hat{A}$, the result will be found to be $\af$. This is because, among all states that assign probability 1 to $[\hat{A} = \alpha]$, $\hat{\Pi}(\hat{A} = \alpha)\psi$ is the closest to $\psi$, as the very notion of  \textit{orthogonal} projection implies. Hence, $\hat{\Pi}(\hat{A} = \alpha)\psi$ is the state in $\hat{\Pi}(\hat{A} = \alpha)(\mathcal{H})$ that best approximates $\psi$ in a mathematically precise sense. 

As we know, the state corresponding to the projected vector $\hat{\Pi}(\hat{A} = \alpha)\psi$ is the projection 
\begin{equation} \begin{split}
    T_{[\hat{A} = \af]}(\ketbra{\psi}) &\doteq \left|\frac{\hat{\Pi}(\hat{A} = \alpha)\psi}{\sqrt{P_{\psi}[\hat{A} = \af]}}\right\rangle \left\langle\frac{\hat{\Pi}(\hat{A} = \alpha)\psi}{\sqrt{P_{\psi}[\hat{A} = \af]}}\right|
    \\
    &=\frac{\hat{\Pi}(\hat{A} = \alpha)\ketbra{\psi}\hat{\Pi}(\hat{A} = \alpha)}{P_{\psi}[\hat{A} = \af]}.
\end{split}\end{equation} 

The straightforward way of extending $\mathcal{S}_{0} \ni \ketbra{\psi} \mapsto\label{eq:pureLudersMapping} T_{[\hat{A} = \af]}(\ketbra{\psi}) \in \mathcal{S}_{0}$ to the set $\mathcal{S}$ of all states is by defining
\begin{equation}
    T_{[\hat{A} = \af]}(\hat{\rho}) \doteq\label{eq:firstLuders} \frac{\hat{\Pi}(\hat{A} = \alpha)\hat{\rho}\hat{\Pi}(\hat{A} = \alpha)}{P_{\hat{\rho}}[\hat{A}=\af]}
\end{equation}
for each state $\hat{\rho}$. This is L\"uders' rule \cite{luders1950upate}, an improvement of von Neumann's collapse postulate \cite{neumann2018foundations}. Let's show that this extension is consistent with our analysis.

To begin with, as one can easily check, $T_{[\hat{A} = \af]}(\hat{\rho})$ ensures that $[\hat{A}=\af]$ is true, i.e., measurement of $\hat{A}$ in the state $T_{[\hat{A} = \af]}(\hat{\rho})$ yields the value $\af$ with probability $1$ (at the theoretical level). Next, let $\hat{\rho} = \sum_{i=1}^{m} p_{i} \ketbra{\psi_{i}}$ be any convex decomposition of the state $\hat{\rho}$ in terms of pure states, with $p_{i} > 0$ for each $i$ and $\sum_{i=1}^{m} p_{i} = 1$. Here, the state $\hat{\rho}$ is traditionally interpreted as asserting that the system lies in the state $\ketbra{\psi_{i}}$ with probability $p_{i}$ \cite{nielsen2000quantum}. Let us consider a sequential measurement on a system prepared in the state $\hat{\rho}$, where we first ask which state $\ketbra{\psi_{i}}$ the system is in, and then ask, for that pure state $\ketbra{\psi_{i}}$, if we measure operator $\hat{A}$ on the system, which eigenvalue we yield. For this state $\hat{\rho}$, the probability $P_{\hat{\rho}}[\rho=\psi_{i},\hat{A} = \af]$ that if we make this sequential measurement, we see first that the system is in the state $\ketbra{\psi_{i}}$ (denoted by $[\rho=\psi_{i}]$) and that, in this pure state $\ketbra{\psi_{i}}$, a measurement of $\hat{A}$ yields the value $\af$, is given by
\begin{equation} \begin{split}
    P_{\hat{\rho}}[\rho=\psi_{i},\hat{A} = \af] &= p_{i} P_{\psi_{i}}[\hat{A} =\af].
\end{split}\end{equation}
For simplicity, we will say that $P_{\hat{\rho}}[\rho=\psi_{i},\hat{A} = \af]$ is the probability of $[\hat{\rho}=\psi_{i}]$ and $[\hat{A}=\af]$ being ``consecutively true'' in the state $\hat{\rho}$. In the same state, $P_{\hat{\rho}}[\hat{A}=\af]$ is the probability that $[\hat{A}=\af]$ is true. The ratio
\begin{equation}
    \frac{P_{\hat{\rho}}[\rho=\psi_{i},\hat{A} = \af]}{P_{\hat{\rho}}[\hat{A}=\af]}
\end{equation}
is thus the probability that $[\hat{\rho}=\psi_{i}]$ and $[\hat{A} =\af]$ are consecutively true, provided that $[\hat{\rho} \in \{\ketbra{\psi_{i}},i=1,\dots,m\} ]$ (i.e., that the system is in one of the states $\ketbra{\psi_{i}},i=1,\dots,m$ ) and $[\hat{A}=\af]$ are satisfied. Now note that
\begin{equation} \begin{split}
    T_{[\hat{A} = \af]}(\hat{\rho}) &=  \sum_{i=1}^{m}p_{i}\frac{\hat{\Pi}(\hat{A} = \alpha)\ketbra{\psi_{i}}\hat{\Pi}(\hat{A} = \alpha)}{P_{\hat{\rho}}[\hat{A}=\af]}
    \\
    &=  \sum_{i=1}^{m}p_{i}\frac{P_{\psi_{i}}[\hat{A} =\af]}{P_{\hat{\rho}}[\hat{A}=\af]}\frac{\hat{\Pi}(\hat{A} = \alpha)\ketbra{\psi_{i}}\hat{\Pi}(\hat{A} = \alpha)}{P_{\psi_{i}}[\hat{A} =\af]}\\
    &= \sum_{i=1}^{m}\frac{p_{i}P_{\psi_{i}}[\hat{A} =\af]}{P_{\hat{\rho}}[\hat{A}=\af]}T_{[\hat{A} = \af]}(\ketbra{\psi_{i}})\\
    &= \sum_{i=1}^{m}\frac{P_{\hat{\rho}}[\rho=\psi_{i},\hat{A} = \af]}{P_{\hat{\rho}}[\hat{A}=\af]}T_{[\hat{A} = \af]}(\ketbra{\psi_{i}}).
\end{split}\end{equation}
Hence, $T_{[\hat{A} = \af]}(\hat{\rho})$ asserts that the system is in one of the states $T_{[\hat{A} = \af]}(\ketbra{\psi_{i}})$, $i=1,\dots,m$, which are pure states ensuring that $[\hat{A}=\af]$ holds. The probability that the system is in the pure state $T_{[\hat{A} = \af]}(\ketbra{\psi_{i}})$ corresponds to the likelihood of $[\hat{\rho}=\psi_{i}]$ and $[\hat{A} =\af]$ being consecutively true, provided that $[\hat{\rho} \in \{\ketbra{\psi_{i}},i=1,\dots,m\} ]$ and $[\hat{A}=\af]$ are satisfied. It shows that our analysis of the state update rule remains valid under L\"uders' extension.

The last step is to accommodate propositions that do not necessarily specify the exact value of the measured quantity --- propositions $[\hat{A} \in \Delta]$ in which $\Delta$ includes more than one element. There is only one consistent way of doing so. Let $\Delta_{1},\dots,\Delta_{m} \subset \sigma(\hat{A})$ be pairwise disjoint sets. Saying that a measurement of $\hat{A}$ will yield a value in $\Delta \doteq \bigcup_{i=1}^{m} \Delta_{i}$ is equivalent to saying that value lying in $\Delta_{i}$ will be obtained for one, and only one, $i=1,\dots,m$. Furthermore, under the evidence that $[\hat{A} \in \Delta]$ is true, the probability that the resulting value lies in $\Delta_{i}$ is given by the conditional probability
\begin{equation}
    P_{\hat{\rho}}^{\hat{A}}(\Delta_{i} \vert \Delta) \doteq\label{eq:standardConditionalProbability} \frac{P_{\hat{\rho}}[\hat{A} \in \Delta_{i} \cap \Delta]}{P_{\hat{\rho}}[\hat{A} \in \Delta]}.
\end{equation}
Hence, saying that the state of the system is $T_{[\hat{A} \in \Delta]}(\hat{\rho})$ is equivalent to saying that the state is $T_{[\hat{A} \in \Delta_{i}]}(\hat{\rho})$ with probability $P_{\hat{\rho}}^{\hat{A}}(\Delta_{i} \vert \Delta)$. This leads to the following consistency condition:
\begin{equation}
    T_{[\hat{A} \in \bigcup_{i=1}^{m}\Delta_{i}]}(\hat{\rho}) =\label{eq:convexUpdate} \sum_{i=1}^{m} P_{\hat{\rho}}^{\hat{A}}(\Delta_{i} \vert \bigcup_{i=1}^{m}\Delta_{i}) T_{[\hat{A} \in \Delta_{i}]}(\hat{\rho}).
\end{equation}
In particular, for any $\Delta \subset \sigma(\hat{A})$ we must have
\begin{equation}
    T_{[\hat{A} \in \Delta]}(\hat{\rho}) =\label{eq:convexUpdatePoints} \sum_{\af \in \Delta} P_{\hat{\rho}}^{\hat{A}}(\{\af\} \vert\Delta) T_{[\hat{A} = \af]}(\hat{\rho}).
\end{equation}
(note that $\Delta = \bigcup_{\af \in \Delta}\{\af\}$). Hence, $T_{[\hat{A} \in \Delta]}(\hat{\rho})$ is fully determined by the definition of $T_{[\hat{A} = \af]}(\hat{\rho})$. This state asserts that the system is in one of the states $T_{[\hat{A} = \af]}(\hat{\rho})$, $\af \in \Delta$. The probability that the system is in the state $T_{[\hat{A} = \af]}(\hat{\rho})$ corresponds to the likelihood of $[\hat{A} = \af]$ being true, provided that $[\hat{A}\in \Delta]$ is satisfied.

It is useful to express the convex combination on the right-hand side of Eq.~\eqref{eq:convexUpdatePoints} using Eq.~\eqref{eq:firstLuders} for $T_{[\hat{A} = \af]}(\hat{\rho})$. We have
\begin{equation} \begin{split}
    T_{[\hat{A} \in \Delta]}(\hat{\rho}) &=\label{eq:LudersFinal} \sum_{\af \in \Delta} \frac{P_{\hat{\rho}}[\hat{A}=\af]}{P_{\hat{\rho}}[\hat{A} \in \Delta]} T_{[\hat{A} = \af]}(\hat{\rho})
    \\
    &=\sum_{\af \in \Delta} \frac{\hat{\Pi}(\hat{A} = \alpha) \hat{\rho} \hat{\Pi}(\hat{A} = \alpha)}{P_{\hat{\rho}}[\hat{A} \in \Delta]}.
\end{split}\end{equation}

We will refer to the mapping $T$ that assigns propositions $[\hat{A} \in \Delta]$ to functions $T_{[\hat{A} \in \Delta]}: \mathcal{S} \ri \mathcal{S}$ through Eqs.~\eqref{eq:firstLuders} and \eqref{eq:convexUpdatePoints} (or equivalently through Eq.~\eqref{eq:LudersFinal}) as the state update mapping, or simply as the L\"uders' rule.

\section{Sequential Measurements, Compatibility, and Joint Distributions}\label{sec:compatibility}

We will denote the distribution of an observable $\hat{B}$ in the state $T_{[\hat{A} \in \Delta]}(\hat{\rho})$ by $P_{\hat{\rho}}[\hat{B} \in (\cdot )|\hat{A} \in \Delta]$. That is, for each $\Sigma \subset \sigma(\hat{B})$ we have
\begin{equation}
    P_{\hat{\rho}}[\hat{B} \in \Sigma|\hat{A} \in \Delta] \equiv P_{T_{[\hat{A} \in \Delta]}(\hat{\rho})}[\hat{B} \in \Sigma].
\end{equation}

This notation is clearly inspired by conditional probability but should not be mistaken for it. The goal is to capture the idea that $P_{T_{[\hat{A} \in \Delta]}(\hat{\rho})}[\hat{B} \in \Sigma]$ represents the probability of $[\hat{B} \in \Sigma]$, provided that $\hat{\rho}$ was updated by $[\hat{A} \in \Delta]$. However, the state $\hat{\rho}$ does not necessarily define a probability measure in a measurable space where $[\hat{A} \in \Delta]$ and $[\hat{B} \in \Sigma]$ can be treated as events, limiting the comparison to a notational tool.

The reason why $[\hat{A} \in \Delta]$ and $[\hat{B} \in \Sigma]$ cannot \textit{a priori} be thought of as events in the same probability space is that $\hat{A}$ and $\hat{B}$ may be incompatible observables, meaning that $\hat{A}\hat{B} \neq \hat{B}\hat{A}$. If $\hat{A}$ and $\hat{B}$ are compatible, however, this assumption holds. This is because compatible observables have well-defined joint probabilities, which in turn assign probabilities to both $[\hat{A} \in \Delta]$ and $[\hat{B} \in \Sigma]$. Let's discuss this in detail.

Let $\hat{A}$ and $\hat{B}$ be (not necessarily compatible) observables, and let $\hat{\rho}$ be a state. For each $\af \in \sigma(\hat{A})$ and $\beta \in \sigma(\hat{B})$, define
\begin{align}
    P_{\hat{\rho}}[\hat{A}=\af,\hat{B}=\beta] &\doteq\label{eq:sequentialDistribution}P_{\hat{\rho}}[\hat{A}=\af]P_{\hat{\rho}}[\hat{B}=\beta | \hat{A} = \af]
    \\
    &=\label{eq:sequentialDistributionQuantum} \Exp{\hat{\Pi}(\hat{A}=\alpha)\hat{\Pi}(\hat{B}=\beta)\hat{\Pi}(\hat{A}=\alpha)}
\end{align}
(note that $\{\hat{\Pi}(\hat{A}=\alpha)\hat{\Pi}(\hat{B} = \beta)\hat{\Pi}(\hat{A}=\alpha): (\af,\beta) \in \sigma(\hat{A})\times\sigma(\hat{B})\}$ is a POVM). $P_{\hat{\rho}}[\hat{A}=\af,\hat{B}=\beta]$ can be thought of as the (theoretically constructed) probability of obtaining values $\af$ and $\beta$ by measuring $\hat{A}$ and $\hat{B}$ in sequence. Following the same terminology as above, we say that $P_{\hat{\rho}}[\hat{A}=\af,\hat{B}=\beta]$ is the probability of $[\hat{A}=\af]$ and $[\hat{B}=\beta]$ being ``consecutively true''.

It follows from Eqs.~(\ref{eq:quantumProbabilitySet}), (\ref{eq:firstLuders}) and (\ref{eq:convexUpdate}) that Eq.~(\ref{eq:sequentialDistribution}) defines a probability distribution $p_{\hat{\rho}}[\hat{A} = \cdot \ , \hat{B}=\cdot \ ]$ on $\sigma(\hat{A}) \times \sigma(\hat{B})$. Let $P_{\hat{\rho}}[(\hat{A},\hat{B}) \in \ \cdot \  ]$ be the associated probability measure. That is, for each $\Gamma \subset \sigma(\hat{A}) \times \sigma(\hat{B})$ we have
\begin{equation}\label{eq:sequentialMeasure}
    P_{\hat{\rho}}[(\hat{A},\hat{B}) \in \Gamma] \doteq \sum_{(\af,\beta) \in \Gamma} P_{\hat{\rho}}[\hat{A}=\af,\hat{B}=\beta].
\end{equation}
For clarity, in the particular case where $\Gamma = \Delta \times \Sigma $ for some pair $\Delta \subset \sigma(\hat{A})$ and $\Sigma \subset \sigma(\hat{B})$, we write
\begin{equation}
    P_{\hat{\rho}}[\hat{A} \in \Delta,\hat{B} \in \Sigma]  \equiv P_{\hat{\rho}}[(\hat{A},\hat{B}) \in \Delta \times \Sigma].
\end{equation}

Eqs.~(\ref{eq:quantumProbabilitySet}), (\ref{eq:firstLuders}) and (\ref{eq:convexUpdate}) ensure that, for each $\Delta \subset \sigma(\hat{A})$ and $\Sigma \subset \sigma(\hat{B})$,
\begin{equation}
    P_{\hat{\rho}}[\hat{A} \in \Delta,\hat{B} \in \Sigma] = \label{eq:sequentialtMeasure} P_{\hat{\rho}}[\hat{A} \in \Delta]P_{\hat{\rho}}[\hat{B} \in \Sigma | \hat{A} \in \Delta].
\end{equation}
Eq.~(\ref{eq:sequentialtMeasure}) can be thought of as the (theoretically constructed) probability of obtaining values in $\Delta$ and $\Sigma$ by measuring $\hat{A}$ and $\hat{B}$ in sequence, as our reading of Eq.~(\ref{eq:sequentialDistribution}) demands. To use the terminology of Section \ref{sec:stateUpdate}, this is the probability of $[\hat{A}\in \Delta]$ and $[\hat{B}\in \Sigma]$ being consecutively true. 

This construction can be canonically extended to include any finite sequence of observables $\hat{A}_{1},\dots,\hat{A}_{m}$. For any state $\hat{\rho}$ and sets $\Delta_{i} \subset \sigma(\hat{A}_{i})$, $i=1,\dots,m$, we define
\begin{equation}
    P_{\hat{\rho}}[\hat{A}_{1} \in \Delta_{1},\dots,\hat{A}_{m} \in \Delta_{m}] \doteq\label{eq:finiteSequentialMeasure} \prod_{i=1}^{m} P_{\hat{\rho}}[\hat{A}_{i} \in \Delta_{i}| \hat{A}_{1} \in \Delta_{1},\dots,\hat{A}_{i-1} \in \Delta_{i-1}],
\end{equation}
where $P_{\hat{\rho}}[\hat{A}_{i} \in  \cdot \ | \hat{A}_{1} \in \Delta_{1},\dots,\hat{A}_{i-1} \in \Delta_{i-1}]$ denotes the distribution of $\hat{A}_{i}$ in the state
\begin{equation}
    T_{[\hat{A}_{i-1} \in\Delta_{i-1}]} \circ \dots \circ T_{[\hat{A}_{1} \in \Delta_{1}]}(\hat{\rho}).
\end{equation}
This mapping induces a probability measure $P_{\hat{\rho}}[\hat{A}_{1},\dots,\hat{A}_{m} \in \cdot \ ]$ on (the power set of) the Cartesian product $\prod_{i=1}^{m}\sigma(\hat{A}_{i})$, which represents the prediction of a sequential measurement of $\hat{A}_{1},\dots,\hat{A}_{m}$ in the state $\hat{\rho}$.

In the particular case where $\hat{A}_{1},\dots,\hat{A}_{m}$ are pairwise compatible, $P_{\hat{\rho}}[(\hat{A}_{1},\dots,\hat{A}_{m}) \in \cdot \ ]$ is their standard joint distribution, i.e., for each $\Delta_{i} \subset \sigma(\hat{A}_{i})$, $i=1,\dots,m$, we have
\begin{equation}
    P_{\hat{\rho}}[\hat{A}_{1} \in \Delta_{1},\dots,\hat{A}_{m} \in \Delta_{m}] =\label{eq:standardJoint} \Exp{\prod_{i=1}^{m}\hat{\Pi}(\hat{A}_{i} \in \Delta_{i})}.
\end{equation}
This follows directly from Eq.~(\ref{eq:sequentialDistributionQuantum}) and well-known results of linear algebra.
More importantly, two observables $\hat{A}$ and $\hat{B}$ are compatible if and only if the probability measures $P_{\hat{\rho}}[(\hat{A},\hat{B}) \in  \cdot \ ]$ and $P_{\hat{\rho}}[(\hat{B},\hat{A}) \in \cdot \ ]$ are equal (up to a permutation):

\begin{lemma}[Compatibility]\label{lemma:compatibility} Let $\mathfrak{S}$ be a finite-dimensional quantum system. Two observables $\hat{A}$ and $\hat{B}$ of $\mathfrak{S}$ are compatible if and only if, for any state $\hat{\rho}$ and  sets $\Delta\subset \sigma(\hat{A})$,$\Sigma \subset \sigma(\hat{B})$,
\begin{equation}
    P_{\hat{\rho}}[\hat{A} \in \Delta]P_{\hat{\rho}}[\hat{B} \in \Sigma | \hat{A} \in \Delta] =\label{eq:bayesLemma}P_{\hat{\rho}}[\hat{B} \in \Sigma]P_{\hat{\rho}}[\hat{A} \in \Delta | \hat{B} \in \Sigma]. 
\end{equation}
\end{lemma}
The proof can be found in Appendix~\ref{sec:proofs}. Note that Eq.~(\ref{eq:bayesLemma}) is formally equivalent to Bayes' rule. This result, along with the subsequent discussion, extends to sets of pairwise compatible observables. We limit the presentation to pairs of observables for clarity.

If $\hat{A}$ and $\hat{B}$ are compatible, the predictions of $\hat{A}$ correspond to those of $\hat{A}$ and $\hat{B}$ marginalised over $\hat{B}$, and vice versa. Consequently, the propositions $[\hat{A} \in \Delta]$ and $[\hat{A} \in \Delta] \wedge [\hat{B} \in \sigma(\hat{B})]$ are equally probable for all states. That is, for each $\Delta \subset \sigma(\hat{A})$ and each state $\hat{\rho}$,
\begin{equation}
    P_{\hat{\rho}}[\hat{A} \in \Delta ,\hat{B} \in \sigma(\hat{B})] =\label{eq:nondisturbanceA} P_{\hat{\rho}}[\hat{A} \in \Delta]
\end{equation}
(analogously, $P_{\hat{\rho}}[\hat{A} \in \sigma(\hat{A}), \hat{B} \in \Sigma] = P_{\hat{\rho}}[\hat{B} \in \Sigma]$ for each $\Sigma \subset \sigma(\hat{B})$). This is referred to in the literature as the non-disturbance condition \cite{amaral2018graph}. 

More importantly,  the propositions $[\hat{A} \in \Delta]$ and  $[\hat{B} \in \Sigma]$ update the joint distribution of $\hat{A}$ and $\hat{B}$ by conditioning it on the set of values where they hold. This means that, for any state $\rho$ and any Borel sets $\Delta',\Sigma'$,
\begin{align}
    P_{\rho}[\hat{A} \in \Delta',\hat{B} \in \Sigma'|\hat{A} \in \Delta,\hat{B} \in \Sigma]&=\label{eq:jointConditioningFormalLemma}\frac{P_{\rho}[\hat{A} \in \Delta' \cap \Delta,\hat{B} \in \Sigma \cap\Sigma']}{P_{\rho}[\hat{A} \in \Delta,\hat{B} \in \Sigma]}.
\end{align}
Together with Eq.~\ref{eq:nondisturbanceA}, it implies that
\begin{align}
        P_{\rho}[\hat{B} \in \Sigma | \hat{A} \in \Delta] &=\label{eq:compatibleUpdateB} P_{\rho}[\hat{A} \in \sigma(\hat{A}),\hat{B} \in \Sigma | \hat{A} \in \Delta,\hat{B} \in \sigma(\hat{B})],\\
        P_{\rho}[\hat{A} \in \Delta | \hat{B} \in \Sigma] &=\label{eq:compatibleUpdateA} P_{\rho}[\hat{A} \in \Delta,\hat{B} \in \sigma(\hat{B}) | \hat{A} \in \sigma(\hat{A}),\hat{B} \in \Sigma].
\end{align}

To conclude, it is worth emphasising that, in the particular case where $\hat{B}=g(\hat{A})$ for some function $g$, the proposition $[g(\hat{A}) \in \Delta]$ conditions the distribution of $\hat{A}$ on the set of values $\af \in \sigma(\hat{A})$ in which $[g(\hat{A}) \in \Delta]$ holds, namely the set $g^{-1}(\Delta) =\{\af \in \sigma(\hat{A}):g(\af) \in \Delta\}$. That is, for each $\Sigma \subset \sigma(\hat{A})$ we have
\begin{equation} \begin{split}
    P_{\hat{\rho}}[\hat{A} \in \Sigma|g(\hat{A}) \in \Delta] &=\label{eq:functionConditioning}P_{\rho}[\hat{A} \in \Sigma|\hat{A} \in g^{-1}(\Delta)]
    \\
    &=\frac{P_{\hat{\rho}}[\hat{A} \in \Sigma \cap g^{-1}(\Delta)]}{P_{\hat{\rho}}[\hat{A} \in g^{-1}(\Delta))]}
\end{split}\end{equation}

In summary, there are two equivalent ways to represent propositions involving compatible observables $\hat{A}$ and $\hat{B}$ as events in a shared probability space. First, we can utilise their joint distribution. In this approach, we associate $[\hat{A} \in \Delta]$ and $[\hat{B} \in \Sigma]$ with $[\hat{A} \in \Delta] \wedge [\hat{B} \in \sigma(\hat{B})]$ and $[\hat{A} \in \sigma(\hat{A})] \wedge [\hat{B} \in \Sigma]$, respectively, where $\wedge$ stands for ``and''. Alternatively, we can represent $\hat{A}$ and $\hat{B}$ as random variables $g$ and $h$ in the spectrum of some observable $\hat{C}$ such that $\hat{A} = g(\hat{C})$ and $\hat{B} = h(\hat{C})$ (it is well-known that $A$ and $B$ are compatible if and only if such an observable exists). In this latter case, $[g(\hat{C}) \in \Delta]$ and $[h(\hat{C}) \in \Sigma]$ are transformed into $[\hat{C} \in g^{-1}(\Delta)]$ and $[\hat{C} \in h^{-1}(\Sigma)]$. Both these constructions explicitly fail if $\hat{A}$ and $\hat{B}$ are incompatible: they lack a joint probability distribution and are not functions of a common observable. There is, therefore, a latent tension between the incompatibility and (deterministic) hidden variables, as the latter presuppose a probability space in which all propositions about a system correspond to events.

\section{Ontological Models Framework}\label{sec:omf}

In the ontological models framework \cite{spekkens2005contextuality, harrigan2010epistemic}, quantum systems are modelled in a space $\boldsymbol{\Lambda}$ of ``ontic states''. As in Bell's \cite{bell1964epr,bell1985beables} and Kochen and Specker's \cite{kochen1975problem} hidden variable models, states $\hat{\rho}$ of $\mathfrak{S}$ are taken to to represent ``the probabilities that the system be in different ontic states'' \cite{spekkens2005contextuality}. Therefore, in this framework, each state $\hat{\rho}$ defines a probability measure $\mu_{\hat{\rho}}$ on $\boldsymbol{\Lambda} \equiv (\Lambda,\mathcal{A})$, where $\Lambda$ is the set of ontic states and $\mathcal{A}$ is an appropriate $\sigma$-algebra on $\Lambda$. Given any measurable set $\Omega \subset \Lambda$, $\mu_{\hat{\rho}}(\Omega)$ represents the probability that the system is in an ontic state lying in $\Omega$. Ontic states are thought of as assigning probabilities to values to observables. It means that each $\lambda \in \Lambda$ defines a probability distribution $\kappa_{\lambda}[\hat{A} = \cdot \ ]:\sigma(\hat{A}) \ri [0,1]$ for each observable $\hat{A}$. Analogously to quantum states, $\kappa_{\lambda}[\hat{A} =\af]$ represents the probability that a measurement of $\hat{A}$ in the ontic state $\lambda$ returns the value $\af$ (the probability of $[\hat{A} =\af]$).



We denote by $\kappa_{\lambda}[\hat{A} \in  \cdot \ ]$ the probability measure associated with the distribution $\sigma(\hat{A}) \ni \af \mapsto \kappa_{\lambda}[\hat{A} = \af] \in [0,1]$. That is, for any set $\Delta \subset \sigma(\hat{A})$ we have
\begin{equation}
    \kappa_{\lambda}[\hat{A} \in \Delta] \doteq \sum_{\af \in \Delta}\kappa_{\lambda}[\hat{A} = \af].
\end{equation}
Note that $\kappa_{\lambda}[\hat{A} \in \Delta]$ represents 
the probability of obtaining a value in $\Delta$ in a measurement of $\hat{A}$ given that the system is in the ontic state $\lambda$. As discussed extensively in Section~\ref{sec:stateUpdate}, this is the probability that the proposition $[\hat{A} \in \Delta]$ (``If a measurement is made of $\hat{A}$, the result will be found to lie in $\Delta$'' \cite{doring2011what}) is true, with the collection of all propositions $[\hat{A}=\af'],\af' \in \sigma(\hat{A}),$ forming the underlying sample space.

The advantage of working with the probability measure $\kappa_{\lambda}[\hat{A} \in \cdot \ ]$ rather than its associated distribution is that it does not exclude observables with continuous spectra, thereby paving the way for extending our results to infinite-dimensional systems. While distributions are more commonly used in the literature on contextuality, probability measures and Markov kernels are standard in probability theory \cite{klenke2014probability, cohn2013measure,tao2011measure}.

To standardise our notation, we will write
\begin{equation}
    \mu_{\hat{\rho}}[\lambda \in \Omega] \equiv\label{eq:intuitiveMeasure} \mu_{\hat{\rho}}(\Omega),
\end{equation}
as $\mu_{\hat{\rho}}(\Omega)$ represents the probability that, in the quantum state $\hat{\rho}$, the system's ontic state $\lambda$ lies in the measurable set $\Omega \subset \Lambda$. Put differently, $\mu_{\hat{\rho}}[\lambda \in \Omega]$ is the probability that the proposition $[\lambda \in \Omega]$ (``the system's ontic lies in $\Omega$'') is true in the state $\hat{\rho}$. 

The predictions of the model must match the quantum predictions. Let $\hat{\rho}$ be a state and $\hat{A}$ be an observable of $\mathfrak{S}$. In quantum mechanics, the probability $P_{\hat{\rho}}[\hat{A} \in \Delta]$ that a measurement of $\hat{A}$ in the state $\hat{\rho}$ yields a value lying in the set $\Delta \subset \sigma(\hat{A})$, i.e., the probability that $[\hat{A} \in \Delta]$ is true, is given by Eq.~(\ref{eq:quantumProbabilitySet}). However, if $\kappa_{\lambda}[\hat{A} \in \Delta]$ represents the probability that a measurement of $\hat{A}$ in the ontic state $\lambda$ returns a value lying in $\Delta$ (the probability of $[\hat{A} \in \Delta]$), and if $\mu_{\hat{\rho}}(\Omega)$ is the probability that the system is in a ontic state lying in $\Omega \subset \Lambda$, then necessarily
\begin{equation}
    P_{\hat{\rho}}[\hat{A} \in \Delta] =\label{eq:probabilityOntologicalModelPre} \int_{\Lambda} \kappa_{\lambda}[\hat{A} \in \Delta] \  \mu_{\hat{\rho}}(d\lambda),
\end{equation}
where the integral in the right-hand side is the Lebesgue integral \cite{halmos1974measure,klenke2014probability}. Note that, for this integral to be well-defined, $\Lambda \ni \lambda \mapsto \kappa_{\lambda}[\hat{A} \in \Delta] \in [0,1]$ must be a measurable function, which in turn means that $\kappa_{( \cdot )}[\hat{A} \in \cdot \ ]$ is a Markov kernel \cite{klenke2014probability}.

To summarise, an ontological model for $\mathfrak{S}$ consists of a measurable space $\boldsymbol{\Lambda} \equiv (\Lambda;\mathcal{A})$, a mapping $\Psi$ assigning states $\hat{\rho}$ of $\mathfrak{S}$ to probability measures $\Psi(\hat{\rho}) \equiv \mu_{\hat{\rho}}$ on $\boldsymbol{\Lambda}$, and a mapping $\Phi$ assigning observables $\hat{A}$ of $\mathfrak{S}$ to Markov kernels $\Phi(\hat{A}) \equiv \kappa_{(\cdot)}[ \hat{A} \in \cdot \ ]$. Furthermore, the model $\mathfrak{M} \equiv (\boldsymbol{\Lambda},\Phi,\Psi)$ must satisfy Eq.~(\ref{eq:probabilityOntologicalModelPre}) for each state $\hat{\rho}$, each observable $\hat{A}$ and each set $\Delta \subset \sigma(\hat{A})$. To accommodate POVMs, one simply replace the mapping $\Phi$ with a function assigning POVMs to Markov kernels, maintaining the rest unchanged. Hereafter, we will denote by $\mathcal{S}(\boldsymbol{\Lambda})$ and $\mathcal{O}(\boldsymbol{\Lambda})$, respectively, the sets of probability measures and Markov Kernels of $\boldsymbol{\Lambda}$. 

\subsection{State Update in Ontological Models}

Let $\mathfrak{M} \equiv (\boldsymbol{\Lambda},\Phi,\Psi)$ be an ontological model for a (finite-dimensional) quantum system $\mathfrak{S}$. To properly model $\mathfrak{S}$ in agreement with the standard postulates of quantum mechanics, $\mathfrak{M}$ must be equipped with a state update mechanism $\tau$ replicating the quantum state update $T$. In the model, states are probability measures, thus $\tau$ is a mapping associating each observable $\hat{A}$ and each set $\Delta \subset \sigma(\hat{A})$, namely each proposition $[\hat{A} \in \Delta]$, to a function $\tau_{[\hat{A} \in \Delta]}: \mathcal{S}(\boldsymbol{\Lambda}) \ri\mathcal{S}(\boldsymbol{\Lambda})$ (recall that $\mathcal{S}(\boldsymbol{\Lambda})$ is the set of probability measures on $\boldsymbol{\Lambda}$). For $\tau$ to replicate the quantum update in the model, the following diagram must commute for every state $\hat{\rho}$ and proposition $[\hat{A} \in \Delta]$:
\begin{center}
        \begin{tikzcd}[column sep=large, arrows={|->}]
        \hat{\rho}\arrow[r,"\Psi"]\arrow[d,"T_{[\hat{A} \in \Delta]}"] & \mu_{\hat{\rho}}\arrow[d,"\tau_{[\hat{A} \in \Delta]}"] \\
        T_{[\hat{A} \in \Delta]}(\hat{\rho}) \arrow[r,"\Psi"] &\mu_{T_{[\hat{A} \in \Delta]}(\hat{\rho})}
        \end{tikzcd}
\end{center}
It means that the probability measure we obtain by updating $\mu_{\hat{\rho}}$ in the model, namely $\tau_{[\hat{A} \in \Delta]}(\mu_{\hat{\rho}})$, is the measure representing the updated state $T_{[\hat{A} \in \Delta]}(\hat{\rho})$. More explicitly,
\begin{equation}
    \Psi(T_{[\hat{A} \in \Delta]}(\hat{\rho})) =\label{eq:embeddingUpdate} \tau_{[\hat{A} \in \Delta]}(\Psi(\hat{\rho})).
\end{equation}

We can construct $\tau$ from the same considerations that led us to Eq.~(\ref{eq:probabilityOntologicalModelPre}). To begin with, recall that the updated state $T_{[\hat{A} \in \Delta]}(\hat{\rho})$ ensures (at the theoretical level) that, if measured, $\hat{A}$ will return a value in $\Delta$. In ontological models, measurements are thought of as ``measurements \textit{of} the ontic state of the system'' \cite{spekkens2005contextuality} which ``might only enable one to infer probabilities for the system to have been in different ontic states'' \cite{spekkens2005contextuality} --- put differently, a measurement ``leads to an update in one’s information about what the ontic state of the system was prior to the procedure being implemented'' \cite{spekkens2005contextuality}. Quantum states, on the other hand, represent ``the probabilities that the system be in different ontic states'' \cite{spekkens2005contextuality}. Hence, in the updated state $T_{[\hat{A} \in \Delta]}(\hat{\rho})$, the probability that the system is in ontic state $\lambda$ is directly proportional to the probability that a measurement of $\hat{A}$ in $\lambda$ returns a value in $\Delta$: the higher the probability $\kappa_{\lambda}[\hat{A} \in \Delta]$, the greater the probability that the system is in this state, because we know that a measurement of $\hat{A}$ in the state $T_{[\hat{A} \in \Delta]}(\hat{\rho})$ will return a value in $\Delta$. Equivalently, the lower the value $\kappa_{\lambda}[\hat{A} \in \Delta]$, the smaller the probability that the system is in this state. It means that, in the state $T_{[\hat{A} \in \Delta]}(\hat{\rho})$, the probability that the system is in an ontic state lying in a measurable set $\Omega \subset \Lambda$ is proportional to $\int_{\Omega}\kappa_{\lambda}[\hat{A} \in \Delta] \ d\mu_{\hat{\rho}}( \lambda)$. Eq.~(\ref{eq:probabilityOntologicalModelPre}) ensures that the normalisation constant is $P_{\hat{\rho}}[\hat{A} \in \Delta]$. Hence, the probability measure representing the updated state $T_{[\hat{A} \in \Delta]}(\hat{\rho})$ is given by
\begin{equation}
    \mu_{T_{[\hat{A} \in \Delta]}(\hat{\rho})}[\lambda \in\Omega] \doteq\label{eq:updateOntologicalModelPre} \int_{\Omega} \frac{\kappa_{\lambda}[\hat{A} \in \Delta]}{P_{\hat{\rho}}[\hat{A} \in \Delta]} \ \mu_{\hat{\rho}}(d\lambda)
\end{equation}
for any measurable set $\Omega \subset \Lambda$. Denoting by $\mu_{\hat{\rho}}( \ \cdot \ |\hat{A} \in \Delta]$ this probability measure, we can write
\begin{equation} \begin{split}
    \mu_{\hat{\rho}}[\lambda \in \Omega|\hat{A} \in \Delta] &=\label{eq:updateOntologicalModelPreSmart} \int_{\Omega} \frac{\kappa_{\lambda}[\hat{A} \in \Delta]}{P_{\hat{\rho}}[\hat{A} \in \Delta]} \ \mu_{\hat{\rho}}(d\lambda)
    \\
    &= \int_{\Omega} \mu_{\hat{\rho}}(d\lambda |\hat{A} \in \Delta]
\end{split}\end{equation}

As in quantum mechanics, Eq.~(\ref{eq:updateOntologicalModelPre}) is, by construction, a mechanism for reconstructing states through propositions about observables. That is, $\tau_{[\hat{A} \in \Delta]}(\mu_{\hat{\rho}})$ reconstructs $\mu_{\hat{\rho}}$ to ensure that the proposition $[\hat{A} \in \Delta]$ holds true. The update state $\tau_{[\hat{A} \in \Delta]}(\hat{\rho})$ is the measure in which $[\hat{A} \in \Delta]$ holds that closely approximates $\mu_{\hat{\rho}}$ (Lemma \ref{lemma:variationDistance} makes this claim mathematically precise in deterministic models).

In summary, the way ontological models are typically interpreted suggests that, to account for the state update, they must be defined as follows.

\begin{definition}[State-updating Ontological model]\label{def:ontologicalModel}
Let $\mathfrak{S}$ be a finite-dimensional quantum system, and let $\mathcal{O}$ and $\mathcal{S}$ be its sets of observables and states, respectively. Let $\mathcal{O}_{S}$ be a non-empty subset of $\mathcal{O}$. An ontological model for $\mathcal{O}_{S}$ consists of a measurable space $\boldsymbol{\Lambda} \equiv (\Lambda,\mathcal{A})$, a mapping $\Psi$ assigning states $\hat{\rho}$ of $\mathfrak{S}$ to probability measures $\Psi(\hat{\rho}) \equiv \mu_{\hat{\rho}}$ on $\boldsymbol{\Lambda}$, and a mapping $\Phi$ assigning observables $\hat{A}$ of $\mathfrak{S}$ to Markov kernels $\Phi(\hat{A}) \equiv \kappa_{(\cdot)}[ \hat{A} \in \cdot \ ]$ such that the following conditions are satisfied. 
\begin{itemize}
    \item[(a)] For any observable $\hat{A} \in \mathcal{O}_{S}$ and any state $\hat{\rho}$, the distribution of $\hat{A}$ w.r.t. $\hat{\rho}$ is given by
    \begin{equation} \begin{split}
    P_{\hat{\rho}}[\hat{A} \in \Delta] &=\label{eq:probabilityOntolgoicalDef}\int_{\Lambda} \kappa_{\lambda}[\hat{A} \in \Delta] \ \mu_{\hat{\rho}}(d\lambda).
    \end{split}\end{equation}
    where the integral in the right-hand side is the Lebesgue integral \cite{halmos1974measure,klenke2014probability}.
     \item[(b)] The state update translates to $\boldsymbol{\Lambda}$ as conditional probability. It means that, for any state $\hat{\rho}$, any observable $\hat{A} \in \mathcal{O}_{S}$ and any  set $\Delta \subset \sigma(\hat{A})$, we have
    \begin{equation}
        \Psi(T_{[\hat{A} \in \Delta]}(\hat{\rho})) =\label{eq:updateOntologicalDef} \tau_{[\hat{A} \in \Delta]}(\Psi(\hat{\rho})),
    \end{equation}
     where $T_{[\hat{A} \in \Delta]}(\hat{\rho}):\mathcal{S} \ri \mathcal{S}$ is given by Lüders' rule (Eq.~(\ref{eq:LudersFinal})) and $\tau_{[\hat{A} \in \Delta]}:\mathcal{S}(\Lambda) \ri \mathcal{S}(\Lambda)$ is determined by Eq.~(\ref{eq:updateOntologicalModelPreSmart}). In particular, for any state $\hat{\rho}$ and any measurable set $\Omega \subset \Lambda$ we have
    \begin{equation}
    \mu_{\hat{\rho}}[\lambda \in \Omega | \hat{A} \in \Delta] \doteq\label{eq:conditionalProbabilityModel} \int_{\Omega}\frac{\kappa_{\lambda}[\hat{A} \in \Delta]}{P_{\hat{\rho}}[\hat{A} \in \Delta]} \ \mu_{\hat{\rho}}(d\lambda),
    \end{equation}
    in which $\mu_{\hat{\rho}}[\lambda \in \Omega | \hat{A} \in \Delta] \equiv \mu_{T_{[\hat{A} \in \Delta]}(\rho)}(\Omega)$. 
\end{itemize}    
\end{definition}

In the particular case where $\mathcal{O}_{S} = \mathcal{O}$, we say that $\mathfrak{M} \equiv (\boldsymbol{\Lambda},\Phi,\Psi)$ is a model for the system $\mathfrak{S}$. We define ontological models for sets of observables, and not for the entire quantum system, for practical reasons. The subscript $S$ stands for scenario, and the set $\mathcal{O}_{S}$ plays the role of a measurement scenario in the contextuality literature.

\subsection{Deterministic Ontological Models}\label{sec:detModels}

An ontological model $\boldsymbol{\Lambda} \equiv (\Lambda;\mathcal{A})$ for a set of observables  $\mathcal{O}_{S}$ is said to be \textit{deterministic} (``for outcomes'' \cite{spekkens2005contextuality}) if the Markov kernels representing observables are deterministic. That is, for each observable $\hat{A} \in \mathcal{O}_{S}$ and each $\af \in \sigma(\hat{A})$, we have
\begin{equation}
    \kappa_{\lambda}[\hat{A} =\af] \in \{0,1\}
\end{equation}
for all $\lambda\in \Lambda$.

Each $\lambda \in \Lambda$ satisfies $\kappa_{\lambda}[\hat{A} =\af] = 1$ for one, and only one, $\af \in \sigma(\hat{A})$, because $\af \mapsto \kappa_{\lambda}[\hat{A} =\af]$ is a probability distribution on $\sigma(\hat{A})$. The set
\begin{equation}
    \Omega(\hat{A} = \af) \doteq \{\lambda \in \Lambda: \kappa_{\lambda}[\hat{A} =\af] =1\}
\end{equation}
consists of all ontic states for which the proposition $[\hat{A} =\af]$ is true. The collection $\Omega(\hat{A} = \af)$, $\af \in \sigma(\hat{A})$, defines a partition of $\Lambda$, meaning they are pairwise disjoint and their union covers the entire set $\Lambda$. Hence, the deterministic Markov kernel $\kappa_{\lambda}[\hat{A} \in \cdot \ ] $ is determined by a measurable function $f_{\hat{A}}$ on $\Lambda$, where each $\Omega(\hat{A} = \af)$ is the pre-image of some value of $f_{\hat{A}}$. The function $f_{\hat{A}}$ of interest assigns ontic states in $\Omega(\hat{A} = \af)$ to the value $\af$, as a measurement of $\hat{A}$ in the ontic state $\lambda$ returns the value $\af$. It means that
\begin{equation}
    \Omega(\hat{A} =\af) = f_{\hat{A}}^{-1}(\{\af\}).
\end{equation}
In particular, it follows from the definition of Lebesgue integral that, for each $\Delta \subset \sigma(\hat{A})$ and each state $\hat{\rho}$,
\begin{equation}
    P_{\hat{\rho}}[\hat{A} \in \Delta] = \mu_{\hat{\rho}}[\lambda \in \Omega(\hat{A}\in \Delta)],
\end{equation}
where
\begin{equation}
    \Omega(\hat{A}\in \Delta) \doteq \bigcup_{\af \in \Delta} \Omega(\hat{A}=\af) = f_{\hat{A}}^{-1}(\Delta).
\end{equation}
That is, $P_{\hat{\rho}}[A \in \Delta]$ is the probability that, in the quantum state $\hat{\rho}$, the system lies in an ontic state in which the proposition $[\hat{A} \in \Delta]$ is true, meaning that $\kappa_{\lambda}[\hat{A} \in \Delta]=1$, or equivalently $f_{\hat{A}}(\lambda) \in \Delta$.

It follows from the definition of Lebesgue integrals of simple functions that, in the particular case of deterministic models, Eq.~(\ref{eq:updateOntologicalModelPre}) reduces to
\begin{equation} \begin{split}
    \mu_{\hat{\rho}}[\lambda \in \Omega|\hat{A} \in \Delta] &\doteq\label{eq:updateOntologicalModelPreDet} \mu_{\hat{\rho}}(\Omega | \Omega(\hat{A} \in \Delta))
    \\
    &=\frac{\mu_{\hat{\rho}}[\lambda \in \Omega \cap \Omega(\hat{A} \in \Delta)]}{P_{\hat{\rho}}[\hat{A} \in \Delta]}.
\end{split}\end{equation}

\begin{definition}[Deterministic state-updating Ontological model]\label{def:detOntologicalModel}
Let $\mathcal{O}_{S}$ be a set of observables in a finite-dimensional quantum system $\mathfrak{S}$. A state updating ontological model $\mathfrak{M}$ for $\mathcal{O}_{S}$ is said to be deterministic if Markov kernels representing observables are deterministic. Equivalently, $\mathfrak{M}$ consists of a measurable space $\boldsymbol{\Lambda} \equiv (\Lambda,\mathcal{\hat{A}})$, a mapping $\Psi$ assigning states $\hat{\rho}$ of $\mathfrak{S}$ to probability measures $\Psi(\hat{\rho}) \equiv \mu_{\hat{\rho}}$ on $\boldsymbol{\Lambda}$, and a mapping $\Theta$ assigning observables $\hat{A}$ of $\mathfrak{S}$ to random variables $\Theta(\hat{A}) \equiv f_{\hat{A}}:\Lambda \ri \sigma(\hat{A})$ such that the following conditions are satisfied. 
\begin{itemize}
    \item[(a)] For any observable $\hat{A} \in \mathcal{O}_{S}$ and any state $\hat{\rho}$, the distribution of $\hat{A}$ w.r.t. $\hat{\rho}$ is given by
    \begin{equation} \begin{split}
    P_{\hat{\rho}}[\hat{A} \in \Delta] &=\label{eq:probabilityOntolgoicalDetDef} \mu_{\hat{\rho}}(\Omega(A \in \Delta)),
    \end{split}\end{equation}
    where $\Omega(A \in \Delta) \equiv f_{\hat{A}}^{-1}(\Delta)$.
     \item[(b)] The state update translates to $\boldsymbol{\Lambda}$ as conditional probability. It means that, for any state $\hat{\rho}$, observable $\hat{A} \in \mathcal{O}_{S}$, and measurable sets $\Delta \subset \sigma(\hat{A})$ and $\Omega \subset \Lambda$, we have
    \begin{equation} \begin{split}
    \mu_{\hat{\rho}}[\lambda \in \Omega | \hat{A} \in \Delta] &\doteq\label{eq:conditionalProbabilityModelDetDef} \mu_{\hat{\rho}}(\Omega | \Omega(\hat{A} \in \Delta))\\
    \\
    &= \frac{\mu_{\hat{\rho}}(\Omega \cap \Omega(\hat{A} \in \Delta))}{P_{\hat{\rho}}[\hat{A} \in \Delta]},
    \end{split}\end{equation}
    in which $\mu_{\hat{\rho}}[\lambda \in \Omega | \hat{A} \in \Delta] \equiv \Psi(T_{\hat{A} \in \Delta}(\rho))(\Omega)$. 
\end{itemize}    
\end{definition}
Note that this is what we referred to as ``state-updating underlying state model'' in Ref.~\cite{Tezzin2025deterministicModels}.

Hereafter, by \textit{stochastic ontological model} we mean an ontological model that is not deterministic. 

The consequences of ``dropping determinism'' from hidden variable descriptions extend beyond simply rejecting ``outcome determinism''. An important difference between deterministic and stochastic models is that, in the former, propositions regarding the same observable $\hat{A}$ are rendered independent (as events) by the system's ontic states, a condition that may be violated in stochastic models. That is, let $\hat{A}$ be an observable and $\Delta,\Delta'$ be subsets of $\sigma(\hat{A})$. In a (not necessarily state updating) stochastic ontological model, an ontic state $\lambda$ may very well satisfy
\begin{equation}
    \kappa_{\lambda}[\hat{A} \in \Delta \cap \Delta'] \neq\label{eq:stochasticDependent} \kappa_{\lambda}[\hat{A} \in \Delta] \kappa_{\lambda}[\hat{A} \in \Delta'],
\end{equation}
because, according to the definition of ontological models, there is no need that the events $[\hat{A} \in \Delta]$ and $[\hat{A} \in \Delta']$ are independent w.r.t. the probability measure $\kappa_{\lambda}[ \hat{A} \in \cdot \ ]$. In a deterministic model, on the other hand, these events are independent w.r.t. all ontic states, as each ontic state assigns probability $1$ to one, and only one, $\af \in \sigma(\hat{A})$. More explicitly, denoting by $\af_{\lambda}$ the (necessarily unique) eigenvalue $\af$ of $\hat{A}$ such that $\kappa_{\lambda}[\hat{A} =\af] = 1$, we have
\begin{equation}
    \kappa_{\lambda}[\hat{A} \in \Delta \cap \Delta'] = 1 \Leftrightarrow \af_{\lambda} \in \Delta \cap \Delta' \Leftrightarrow
    \begin{cases}
        \kappa_{\lambda}[\hat{A} \in \Delta] = 1,\\
        \kappa_{\lambda}[\hat{A} \in \Delta'] = 1,
    \end{cases}
\end{equation}
and since $\kappa_{\lambda}[\hat{A} \in \Sigma] \in \{0,1\}$ for each $\Sigma \subset \sigma(\hat{A})$,
\begin{equation}
    \kappa_{\lambda}[\hat{A} \in \Delta \cap \Delta'] =\label{eq:deterministicIndependent} \kappa_{\lambda}[\hat{A} \in \Delta] \kappa_{\lambda}[\hat{A} \in \Delta'].
\end{equation}
In other words, the Markov kernel representing an observable $\hat{A}$ in a stochastic model may have overlapping components, meaning that an ontic state $\lambda$ can assign a strictly positive probability to two different eigenvalues of $\hat{A}$ (i.e., $\kappa_{\lambda}[\hat{A} = \af] >0$ for two distinct $\af \in \sigma(\hat{A})$).

\section{Main Results}\label{sec:results}

This section presents a detailed analysis of state-updating ontological models. Section~\ref{sec:analyzingModels} demonstrates that ontological models adhering to Def.~\ref{def:ontologicalModel}, especially deterministic ones, satisfy important consistency conditions for hidden-variable descriptions of quantum mechanics, including Kochen-Specker noncontextuality and Bell's locality condition. Section~\ref{sec:incompatibilityObstructs} presents our main result, namely that incompatibility obstructs the existence of these models. 

\subsection{On the Formal Consistency of State-Updating Ontological Models}\label{sec:analyzingModels}

We start with the following Lemma:

\begin{lemma}[State-updating ontological model]\label{lemma:ontologicalModel} Let $\mathfrak{S}$ be a finite-dimensional quantum system, and let $\mathcal{O}_{S}$ be a non-empty set of observables of $\mathfrak{S}$. Let $\mathfrak{M} \equiv (\boldsymbol{\Lambda},\Phi,\Psi)$ be an ontological model for $\mathcal{O}_{S}$ (Def.~\ref{def:ontologicalModel}). The following conditions are satisfied.
\begin{itemize}
    \item[(a)] For any observable $\hat{A}$, any function $g:\sigma(\hat{A}) \ri \R$ and any state $\hat{\rho}$,
    \begin{equation} \begin{split}
        \Exp{g(\hat{A})} &=\label{eq:expectationOfFunctionModel} \int_{\Lambda} \int_{\sigma(\hat{A})} g(\af) \ \kappa_{\lambda}[\hat{A} \in d\af]  \mu_{\hat{\rho}}(d\lambda),
\end{split}\end{equation}
In particular, the expectation value of $\hat{A}$ is given by
    \begin{equation}
        \Exp{\hat{A}} =\label{eq:expectationModel} \int_{\Lambda} \int_{\sigma(\hat{A})} \af\  \kappa_{\lambda}[\hat{A} \in d\af]  \mu_{\hat{\rho}}(d\lambda).
    \end{equation}
    
    \item[(b)] $\mathfrak{M}$ is Kochen-Specker noncontextual (and consequently Bell-local \cite{bell1985beables}). That is, if $\hat{A}_{1},\dots, \hat{A}_{m} \in \mathcal{O}_{S}$ are pairwise compatible observables, for any state $\hat{\rho}$ and any sets $\Delta_{i}\subset \sigma(\hat{A}_{i})$, $i=1,\dots,m$, we have
    \begin{equation} \begin{split}
    P_{\hat{\rho}}[\hat{A}_{1} \in \Delta_{1},\dots,\hat{A}_{m} \in \Delta_{m}] &=\label{eq:bellCond} \int_{\Lambda} \prod_{i=1}^{m}\kappa_{\lambda}[\hat{A}_{i} \in \Delta_{i}]\ \mu_{\hat{\rho}}(d\lambda),
    \end{split}\end{equation}
    
    \item[(c)] If $\hat{A},\hat{B} \in \mathcal{O}_{S}$ are compatible observables, then
    \begin{equation} \begin{split}
        \Exp{\hat{A} + \hat{B}} &= \int_{\Lambda}\int_{\sigma(\hat{A})}\int_{\sigma(\hat{B})}\af+\beta\ \kappa_{\lambda}[\hat{A} \in d\af] \kappa_{\lambda}[\hat{B} \in d\beta]  \mu_{\hat{\rho}}(d\lambda),\\
        \Exp{\hat{A}\hat{B}} &=\label{eq:expectationProduct} \int_{\Lambda}\int_{\sigma(\hat{A})}\int_{\sigma(\hat{B})}\af\beta \ \kappa_{\lambda}[\hat{A} \in d\af] \kappa_{\lambda}[\hat{B} \in d\beta]  \mu_{\hat{\rho}}(d\lambda).
    \end{split}\end{equation}
\end{itemize}
\end{lemma}
Proof of this can be found in the Appendix~\ref{sec:proofs}.

In deterministic models, these results can be expressed in terms of random variables rather than Markov Kernels (see Appendix \ref{sec:proofs} for details):

\begin{corollary}[Deterministic state-updating ontological model]\label{cor:DeterministicModel} Let $\mathfrak{S}$ be a finite-dimensional quantum system, and let $\mathcal{O}_{S}$ be a non-empty set of observables of $\mathfrak{S}$. Let $\mathfrak{M} \equiv (\boldsymbol{\Lambda},\Phi,\Psi)$ be a deterministic ontological model for $\mathcal{O}_{S}$. The following conditions are satisfied.
\begin{itemize}
    \item[(a)] The expectation of any observable $\hat{A} \in \mathcal{O}_{S}$ is the expected value of the random variable $f_{\hat{A}} \equiv \Phi(\hat{A})$. That is, given any state $\hat{\rho}$, we have
    \begin{equation}
        \Exp{\hat{A}} = \int_{\Lambda} f_{\hat{A}}(\lambda) \ \mu_{\hat{\rho}}(d\lambda).
    \end{equation}
    \item[(b)] $\mathfrak{M}$ is noncontextual. That is, if $\hat{A}_{1},\dots, \hat{A}_{m} \in \mathcal{O}_{S}$ are pairwise compatible observables, for any state $\hat{\rho}$ and any Borel sets $\Delta_{1},\dots,\Delta_{m}$ we have
    \begin{equation} \begin{split}
    P_{\hat{\rho}}[\hat{A}_{1} \in \Delta_{1},\dots,\hat{A}_{m} \in \Delta_{m}] &= \int_{\Lambda} \prod_{i=1}^{m}\kappa_{\lambda}[\hat{A}_{i} \in \Delta_{i}]\ \mu_{\hat{\rho}}(d\lambda)
    \\
    &=\label{eq:bellCondDet}\mu_{\hat{\rho}}(\bigcap_{i=1}^{m} f_{\hat{A}_{i}}^{-1}(\Delta_{i})).
    \end{split}\end{equation}
    
    \item[(c)] If  $\hat{A},\hat{B} \in \mathcal{O}_{S}$ are compatible observables, we have
    \begin{equation} \begin{split}
        \Exp{\hat{A} + \hat{B}} &= \int_{\Lambda} f_{\hat{A}}(\lambda) + f_{\hat{B}}(\lambda) \ \mu_{\hat{\rho}}(d\lambda),\\
        \Exp{\hat{A}\hat{B}} &= \int_{\Lambda} f_{\hat{A}}(\lambda) \cdot f_{\hat{B}}(\lambda)\ \mu_{\hat{\rho}}(d\lambda).
    \end{split}\end{equation}
\end{itemize}
\end{corollary}

The following lemma demonstrates that deterministic models meet key consistency conditions required for any state update mechanism that works as we propose. In stochastic models, this is generally false, as these conditions remain valid only when $\tau$ is restricted to the set of measures corresponding to quantum states. This is because, as discussed in Section~\ref{sec:detModels}, Markov kernels representing observables may have overlapping components, meaning that an ontic state $\lambda$ can assign a strictly positive probability $\kappa_{\lambda}[\hat{A} = \af]$ to two different eigenvalues $\af$ of $\hat{A}$. Such overlaps prevent us from showing that propositions regarding the same observable are independent (as events) w.r.t. ontic states, which means that an ontic state $\lambda$ may  satisfy Eq.~(\ref{eq:stochasticDependent}). To solve this problem, one might represent quantum observables with non-overlapping Markov Kernels, where each ontic state $\lambda$ satisfies $\kappa_{\lambda}[\hat{A}=\af]>0$ for only one $\af \in \sigma(\hat{A})$. However, such model is necessarily deterministic, because the mapping $\sigma(\hat{A}) \ni \af \mapsto \kappa_{\lambda}[\hat{A} = \af] \in [0,1]$ is a probability distribution and, as such, must sum to one. That is, we have $\kappa_{\lambda}[\hat{A}=\af]>0$ for only one $\af \in \sigma(\hat{A})$ if and only if $\kappa_{\lambda}[\hat{A} = \af] = 1$.

\begin{lemma}[State update in deterministic models]\label{lemma:deterministicUpdate}  Let $\mathfrak{S}$ be a finite-dimensional quantum system, and let $\mathcal{O}_{S}$ be a non-empty set of observables of $\mathfrak{S}$. Let $\mathfrak{M} \equiv (\boldsymbol{\Lambda},\Theta,\Psi)$ be a deterministic state-updating ontological model for $\mathcal{O}_{S}$ (Def.~\ref{def:detOntologicalModel}). For any observable $\hat{A} \in \mathcal{O}_{S}$ and any measure $\mu \in \mathcal{S}(\Lambda)$, the following conditions are satisfied.
\begin{itemize}   
    \item[(a)] For any sets $\Delta,\Delta' \subset \sigma(\hat{A})$,
    \begin{equation} \begin{split} 
        P_{\mu}[\hat{A} \in \Delta' | \hat{A} \in \Delta] &=\label{eq:selfConsistencyDist}\frac{P_{\mu}[\hat{A}\in \Delta \cap \Delta']}{P_{\mu}[\hat{A} \in \Delta]}
    \end{split}\end{equation}
    where, for each $\Sigma \subset \sigma(\hat{A})$,  $P_{\mu}(\cdot \ |\hat{A} \in \Sigma] \equiv \tau_{[\hat{A} \in \Sigma]}(\mu)(\ \cdot \ )$.
    \item[(b)]  For any sets $\Delta,\Delta' \subset \sigma(\hat{A})$,
    \begin{equation} \begin{split}
        (\tau_{[\hat{A} \in \Delta']} \circ \tau_{[\hat{A} \in\Delta]})(\mu) &= \tau_{[\hat{A} \in \Delta' \cap \Delta]}(\mu).
    \end{split}\end{equation}
    
    \item[(c)] If $\Delta_{i} \subset \sigma(\hat{A})$, $i=1,\dots,m$, are pairwise disjoint, for any probability measure $\mu$ on $\boldsymbol{\Lambda}$ we have
    \begin{equation}
        \tau_{[\hat{A} \in \bigcup_{i=1}^{m}\Delta_{i}]}(\mu) =\label{eq:convexUpdateDetLemma} \sum_{i=1}^{m} P_{\mu}[\hat{A} \in \Delta_{i}|\hat{A} \in \cup_{i=1}^{m}\Delta_{i}] \tau_{[\hat{A} \in \Delta_{i}]}(\mu).
    \end{equation}
\end{itemize}
\end{lemma}
Item (a) is valid by definition. Item (b) is a well-known and easy to check feature of conditional probability. Item (c) is the famous summation formula \cite{klenke2014probability}.

It is important to note that, in continuous systems, Eq.~(\ref{eq:selfConsistencyDist}) is virtually sufficient to rule out any update that differ from conditional probability. In fact, since measurements in ontological models are  thought of as ``measurements \textit{of} the ontic state of the system'' which ``might only enable one to infer probabilities for the system to have been in different ontic states'' \cite{spekkens2005contextuality}, it is certainly reasonable to suppose that there exists a (possibly multi-valued) observable $\hat{A}$ that does exactly this, i.e., measuring $\hat{A}$ simply reveals in which ontic state the system exists (e.g., in a system of $N$ particles in classical mechanics, $\hat{A}$ is the identity function $\lambda \mapsto \lambda$, which  merely specify all particle's positions and momenta). Given any measurable set $\Omega \subset \Lambda$, there exists a set of values $\Delta$ such that $\Omega = \Omega(\hat{A} \in \Delta)$. Let $\Omega,\Omega' \subset \Lambda$ be measurable sets, and let $\Delta,\Delta'$ be such that $\Omega = \Omega(\hat{A} \in \Delta)$, $\Omega' = \Omega(\hat{A} \in \Delta)$. Let $\tau$ be the state update mechanism of the model, and suppose it respects Eq.~(\ref{eq:selfConsistencyDist}). Then, for any quantum state $\hat{\rho}$,
\begin{equation}
    \tau_{[\hat{A} \in \Delta]}(\mu_{\hat{\rho}})(\Omega') = P_{\rho}[\hat{A} \in \Delta'|\hat{A} \in \Delta] = \frac{P_{\hat{\rho}}[\hat{A} \in \Delta \cap \Delta']}{P_{\hat{\rho}}[\hat{A} \in \Delta]} = \mu_{\hat{\rho}}(\Omega' | \Omega).
\end{equation}
In models where the set of ontic states is finite, the assumption of continuity can be dropped. Therefore, for models where the set of ontic states is finite, outcome repeatability (at the quantum mechanical level) is essentially sufficient to rule out violations of conditional probability. Given \cite{Schmid2018Noncontextuality} shows that, when modelling a finite-dimensional system, one can always assume the space of ontic states is finite, it seems outcome repeatability alone is enough to rule out violations of conditional probability for most systems.

Only deterministic ontological models update all probability measures in accordance with the consistency conditions we identified for the state update rule. Fortunately, there is no loss of generality in focusing on deterministic models, because Kolmogorov's extension theorem (Theorem \ref{thm:kolmogorov}) guarantees the existence of a \textit{deterministic} ontological model whenever a state-updating ontological model can be constructed. Unfortunately, it is a straightforward consequence of Lemma~\ref{lemma:compatibility} and Bayes' rule that such models exist only in the trivial case where all observables under consideration are pairwise compatible.

\subsection{Incompatibility Obstructs State-updating Ontological Models}\label{sec:incompatibilityObstructs}

According to Def.~\ref{def:ontologicalModel}, ontological models are endowed with a reasonable state update mechanism and satisfy important consistency conditions for hidden variable descriptions, as Lemma~\ref{lemma:ontologicalModel} and its corollary show. Unfortunately, the existence of incompatible observables in quantum systems rules out these models. This is because the quantum predictions of sequential measurements of incompatible observables are order-dependent, which sharply contrast with the order-independence of sequential probabilities dictated by conditional probability, as captured by Bayes's rule. Let's show it in detail.

It follows from Lemma~\ref{lemma:compatibility} that incompatibility obstructs the existence of ontological models. In fact, if an ontological model for a set of observables $\mathcal{O}_{S}$ exists, Eq.~(\ref{eq:bayesLemma}) is naturally satisfied for any pair $\hat{A},\hat{B} \in \mathcal{O}_{S}$. This is because, for any pair  $\hat{A},\hat{B} \in \mathcal{O}_{S}$, any state $\hat{\rho}$ and any sets $\Delta,\Sigma$,

\begin{equation}
\begin{split}
    P_{\hat{\rho}}[\hat{A} \in \Delta,\hat{B} \in \Sigma] &= P_{\hat{\rho}}[\hat{A} \in \Delta]P_{\hat{\rho}}[\hat{B} \in \Sigma | \hat{A} \in \Delta]
    \\
    &= P_{\hat{\rho}}[\hat{A} \in \Delta]\int_{\Lambda} \kappa_{\lambda}[\hat{B} \in \Sigma] \ \mu_{\hat{\rho}}(d\lambda | \hat{A} \in \Delta]
    \\
    &=P_{\hat{\rho}}[\hat{A} \in \Delta] \int_{\Lambda} \kappa_{\lambda}[\hat{B} \in \Sigma]\frac{\kappa_{\lambda}[\hat{A} \in \Delta]}{P_{\hat{\rho}}[\hat{A} \in \Delta]} \ \mu_{\hat{\rho}}(d\lambda)
    \\
    &=\int_{\Lambda} \kappa_{\lambda}[\hat{A} \in \Delta]\kappa_{\lambda}[\hat{B} \in \Sigma] \ \mu_{\hat{\rho}}(d\lambda)
    \\
    &=P_{\hat{\rho}}[\hat{B} \in \Sigma]P_{\hat{\rho}}[\hat{A} \in \Delta | \hat{B} \in \Sigma]
    \\
    &=P_{\hat{\rho}}[\hat{B} \in \Sigma,\hat{A} \in \Delta]
\end{split}
\end{equation}
That is, if a model for $\mathcal{O}_{S}$ exists, then the observables in $\mathcal{O}_{S}$ are pairwise compatible. On the other hand, the Kolmogorov extension theorem (Theorem \ref{thm:kolmogorov}) ensures that a \textit{deterministic} ontological models can be constructed for any set of pairwise compatible observables (see Appendix~\ref{sec:proofs} for details). Therefore, the existence of an ontological model for a given scenario is equivalent to assuming that all observables in that scenario are pairwise compatible. We have the following proposition.

\begin{proposition}[Compatibility and ontic states]\label{prop:impossibility} Let $\mathfrak{S}$ be a finite-dimensional quantum system, and let $\mathcal{O}_{S}$ be a non-empty set of observables in $\mathfrak{S}$. Then the following claims are equivalent:
\begin{itemize}
    \item[(a)] $\mathcal{O}_{S}$ admits a (state-updating) ontological model.
    \item[(b)] $\mathcal{O}_{S}$ admits a deterministic (state-updating) ontological model.
    \item[(c)] The observables in $\mathcal{O}_{S}$ are pairwise compatible.
\end{itemize}
\end{proposition}
$\hfill\blacksquare$

Since incompatible observables are always present in quantum systems, the following Corollary immediately follows.

\begin{corollary}[Incompatibility obstructs state-updating ontological models]\label{cor:impossibility} No finite-dimensional quantum system admits a state-updating ontological model.
\end{corollary}

\section{Discussion}\label{sec:disc}

Under our counterfactual account of the quantum rule, and if we are right that such an account translates to ontological models as per Def.~\ref{def:ontologicalModel}, Proposition \ref{prop:impossibility} provides a straightforward method to rule out ontological models of the canonical formulation of quantum mechanics (i.e., directly including the collapse postulate \cite{bongaarts2015quantum,nielsen2000quantum,Hance2022MeasProb}). The proof establishes a direct connection between compatibility and ontic states, demonstrating that incompatibility is sufficient to obstruct their existence, thereby extending the findings of \citet{fine1982commutingObservables,malley2004commuteSimultaneously,malley2005localRealism}. Since our proof imposes no consistency condition on the representation of observables (such as Kochen and Specker's functional relations \cite{kochen1975problem} or Spekkens' ontological identity of empirical indiscernibles \cite{spekkens2019Leibniz}) the result is independent of considerations about either generalised or standard (Kochen-Specker) contextuality, or determinism.

Our argument consists of the following steps:
\begin{itemize}
    \item[(a)] Section~\ref{sec:stateUpdate} establishes a counterfactual account of the  quantum state update according to which L\"uders' rule is a theoretical mechanism for reconstructing states to ensure the validity of counterfactual propositions about future (i.e., subsequent) measurements of physical quantities.
    \item[(b)] Section~\ref{sec:omf} argues that, due to the aforementioned analysis of the quantum state update, ontological models must update states via conditional probability.
    \item[(c)] Section~\ref{sec:analyzingModels} shows that items (a) and (b) above are consistent by demonstrating that state-updating ontological models meet key consistency requirements for hidden variable representations of quantum mechanics.
    \item[(d)] Section~\ref{sec:incompatibilityObstructs} shows that this consistent representation exists only in the trivial case where all observables considered are compatible. In particular, no quantum system admits a state-updating ontological model.
\end{itemize}
To circumvent Corollary~\ref{cor:impossibility}, one must argue that at least one of these steps is incorrect. Items (c) and (d) are mathematical theorems based on assumptions either in (a) or (b), or in the ontological models framework, and so cannot be disputed without also disputing one or more of these other assumptions. We address item (b) in Section~\ref{sect:howexistingmodelsupdatethen}. Sections~\ref{sec:stateUpdateDiscussion} and \ref{sec:contrastingPictures} focus on item (a). Section~\ref{sec:concludingRemarks} presents our concluding remarks.

\subsection{Considerations About the Quantum State Update}\label{sec:stateUpdateDiscussion}

One of the oldest --- and most fascinating --- conceptual challenges in quantum mechanics is the measurement problem \cite{norsen2017foundations,landsman2017foundations,maudlin1995threeProblems,Hance2022MeasProb,muller2023six}. At least six distinct questions are commonly referred to as ``the measurement problem'' \cite{muller2023six}, while some argue that no such problem exists at all \cite{mermin2022noProblem}. Collectively, these discussions lead many to assume that, in the quantum formalism, the theoretically constructed state update inherently reflects the physical collapse of a system's state following an interaction with a measuring apparatus. That is, rather than updated by propositions in the first conditional (``\textit{if} a measurement is made of $A$, the result \textit{will be} found to lie in $\Delta$''), quantum states are often thought to be updated --- even at the theoretical level --- by propositions in the simple past (``a measurement of $A$ \textit{was} made and a value lying in $\Delta$ \textit{was} obtained''). If this is correct, there is no reason for ontological models to respect conditional probability. Here, we defend our counterfactual account, arguing that this linguistic shift is a misconception.

First, when we say that L\"uders' rule reconstructs states to ensure the validity of counterfactual propositions (if a measurement is made...), we are not advocating for a specific interpretation of the state update mechanism; rather, we are merely describing what the update mechanism is from a mathematical standpoint. As shown in Section~\ref{sec:stateUpdate}, the updated state $T_{[A \in \Delta]}(\rho)$, is, by construction, an optimal reconstruction of $\rho$ designed to ensure that, if a measurement is made of $\hat{A}$, the result will be found to lie in $\Delta$. It is natural for it to be so, as quantum states are (at the least) tools for making predictions about systems' future behaviour. Since states assign probabilities to such counterfactual propositions (i.e., to statements about subsequent measurements), we can naturally reconstruct states to ensure their validity. 

Consider now the simple past proposition ``a measurement of $\hat{A}$ was made, and a value lying in $\Delta$ was obtained'', which we denote by $|\hat{A} \in \Delta|$. There is nothing in the quantum formalism connecting the theoretically constructed state $T_{[\hat{A} \in \Delta]}(\rho)$ with such a proposition. As in any physical theory, propositions about the past cannot be validated or dismissed by quantum states, since states assign probabilities only to subsequent measurements\footnote{Unless we are considering a quantum retrodiction scenario---though even here, the way quantum states assign probabilities is far closer to the form described by propositions in the first conditional (just with the arrow of time reversed) than it is to those in the simple past.}. Since the quantity $\hat{A}$ assumes a value $\af$ immediately after being measured (this is precisely what the measurement informs us, up the the region $\Delta$ wherein this value lies), and since a measurement of a physical quantity with a definite value must simply reveal, up to the limited precision encapsulated by $\Delta$, this value to the observer, $T_{[\hat{A} \in \Delta]}(\hat{\rho})$ serves as the ideal representation of the system's state when $\hat{\rho}$ is prepared, $\hat{A}$ is measured, and a value lying in $\Delta$ is obtained. However, this simply means that the experimental post-measurement state is one instance where the system is ideally described by the theoretical construction $T_{[\hat{A} \in \Delta]}(\rho)$. It by no means indicates that $T_{[\hat{A} \in \Delta]}(\rho)$ must necessarily be thought of a state where the proposition $|\hat{A} =\af|$ (the simple past version) is true. That is, interpreting $T_{[\hat{A} \in \Delta]}(\hat{\rho})$ as a post-measurement state is circumstantially possible, but the theoretical formalism by no means suggests that $|\hat{A} \in \Delta|$ is essential for characterising $T_{[\hat{A} \in \Delta]}(\hat{\rho})$, i.e., that the update $\hat{\rho} \mapsto T_{[\hat{A} \in \Delta]}(\hat{\rho})$ is characterised by physical disturbances inherently associated with interactions with a measuring apparatus.

There are many features of the quantum state update that are left without explanation under the post-measurement (past-tense) interpretation of updated states. Repeatability is the most straightforward. A less-discussed  point is that, if $\hat{C}=g(\hat{A})$, the proposition $[g(\hat{A})\in \Delta]$ conditions the distribution of $\hat{A}$ on the set of values in which this proposition holds, namely the set of values $\af \in \sigma(\hat{A})$ such that $g(\af) \in \Delta$ (see Eq.~\eqref{eq:functionConditioning}). In particular, if $\hat{C}=g(\hat{A})=h(\hat{B})$, with $\hat{A}$ and $\hat{B}$ incompatible, the distributions of both $\hat{A}$ and $\hat{B}$ are conditioned on the subsets of $\sigma(\hat{A})$ and $\sigma(\hat{B})$, respectively, in which $[\hat{C} \in \Delta]$ is satisfied, namely $g^{-1}(\Delta)$ and $h^{-1}(\Delta)$. More broadly, $[\hat{C} \in \Delta]$ conditions the distribution of $\hat{D}$ whenever $\hat{C}$ and $\hat{D}$ are compatible, as shown in Section.~\ref{sec:compatibility}. If $\hat{C}$ and $\hat{D}$ are incompatible, it is not possible to use conditional probability because propositions concerning them cannot be represented in the same probability space.

To mention one more difficulty, if the post-measurement interpretation is right, isolated systems would never have a well-defined state (at least not for a straightforward reason), because all pure states are updated states. To illustrate, consider an hydrogen atom in a bound state $\psi_{n,l,m}$, in which the electron has a well-defined energy and is bound to the proton \cite{hall2013quantum}. This state satisfies
\begin{equation}
    \psi_{n,l,m} = (T_{[\hat{L_{3}}=\theta_{m}]} \circ T_{[\hat{L^{2}}=\phi_{l}]} \circ T_{[\hat{H}=E_{n}]})(\hat{\rho}),
\end{equation}
where $\hat{\rho}$ is any state and $\circ$ stands for function composition. $\hat{H}$, $\hat{L^{2}}$ and $\hat{L_{3}}$ are observables corresponding to energy, magnitude of angular momentum, and the projection of angular momentum in the $z$-axis, whereas $E_{n}$, $\psi_{l}$, $\theta_{m}$ are values corresponding to the quantum numbers $n,l,m$ \cite{hall2013quantum}. By the post-measurement interpretation, the state $\psi_{n,l,m}$ can only be assumed by an atom if an agent measures $\hat{H}$, $\hat{L}^{2}$ and $\hat{L}_{3}$ in sequence. This perspective is challenging to reconcile with the stability of matter, which implies that hydrogen atoms out there in the universe --- which have never interacted with apparatuses --- can and do exist in bound states. If one refrains from adopting such view, on the other hand, it is uncontroversial to state that distant atoms lie in bound states. The state $\psi_{n,l,m}$ is characterised by the truth of the propositions $[\hat{H} = E_{n}]$, $[\hat{L}^{2} = \phi_{l}]$, and $\hat{[L}_{3} = \theta_{m}]$ in the system in question, regardless of how this configuration was achieved; in particular, regardless of whether an interaction with a measurement apparatus occurred or not. This is precisely how the bound state $\psi_{n,l,m}$ is traditionally interpreted: we say that, for whatever reason, and regardless of whether a human being is aware of that or not, the electron has energy $E_{n}$, magnitude of angular momentum $\phi_{l}$, and component $z$ of angular momentum $\theta_{m}$.

In summary, in its explicit mathematical form, the quantum state update is a mechanism for reconstructing states based on propositions about the system's future predictions (i.e., counterfactual propositions). Asserting that this update inherently reflects the physical disturbance resulting from a past interaction with a measuring apparatus, i.e., saying the reconstruction mechanism is grounded on propositions in simple past tense, is a claim that, in our view, lacks theoretical support within the standard quantum formalism.  

To conclude, we would like to emphasise that we are not denying the existence of a measurement problem in quantum mechanics. The question of how, \textit{in a real experiment}, a superposition of eigenstates of the measured quantity collapses to one of these states --- and whether such a jump can be described by quantum mechanics --- is indeed rich and fascinating. However, we believe this issue pertains to the experimental collapse and should not affect our understanding of the theoretical mechanism of state reconstruction. Theoretically, the state update mechanism depends solely on predictions about the system's future behaviour. Connections with post-measurement collapses are merely circumstantial and, as such, cannot determine this theoretical construction.

\subsection{Contrasting Pictures of Quantum Systems: OMF vs. Hilbert}\label{sec:contrastingPictures}

Our analysis of the quantum state update places Lüders' rule and conditional probability on the same footing: both serve as mechanisms for reconstructing states to ensure the validity of certain propositions about future (i.e., subsequent) measurements of physical quantities. The only difference is that the former operates within the quantum formalism of Hilbert spaces, whereas the latter relies on a probability space shared by all physical quantities altogether. This section examines the distinctions between these two frameworks and clarifies why the former permits order-dependent predictions in sequential measurements, whereas the latter does not.

Proposition~\ref{prop:impossibility} shows that, at least under our view on the state update, there is no loss of generality in working with deterministic models. In the deterministic state-updating ontological model $\mathfrak{M} \equiv(\boldsymbol{\Lambda},\Phi,\Psi)$ for a set of observables $\mathcal{O}_{S}$ we constructed when proving Proposition~\ref{prop:impossibility}, $\boldsymbol{\Lambda}$ consists of the Cartesian product $\Lambda \doteq \prod_{A \in \mathcal{O}_{S}}\sigma(\hat{A})$ equipped with the product $\sigma$-algebra $\Lambda \doteq \prod_{A \in \mathcal{O}_{S}}\mathfrak{P}(\sigma(\hat{A}))$, where $\mathfrak{P}(\sigma(\hat{A}))$ is the power set of $\sigma(\hat{A})$. There is thus no loss of generality in assuming that deterministic ontological models are equipped with a Hausdorff topology \cite{tao2011measure}, as we do in Lemma~\ref{lemma:variationDistance}. 

In a quantum system $\mathcal{H}$, a proposition $[\hat{A} = \af]$ (``if a measurement of $\hat{A}$ is made, the result will be found to be $\af$'') is represented by the set $\hat{\Pi}(\hat{A}=\af)(\mathcal{H})$ of pure states (here treated as unit vectors) in which this proposition is true, in that $\psi \in \hat{\Pi}(\hat{A}=\af)(\mathcal{H})$ if and only if $P_{\psi}[\hat{A} =\af] =1$. Projecting a state $\psi \in \mathcal{H}$ consists of selecting the state in which $[\hat{A} = \af]$ holds that better approximates $\psi$, as it is the unique element of $\hat{\Pi}(\hat{A}=\af)(\mathcal{H})$ that minimises the distance from $\psi$ (Eq.~\eqref{eq:minimizeDistance}). Hence, projecting $\psi$ orthogonally onto $\hat{\Pi}(\hat{A}=\af)(\mathcal{H})$ is an optimal way of reconstructing $\psi$ to ensure that, in the reconstructed state, $[\hat{A} = \af]$ is true. To accommodate mixed states $\hat{\rho}$, one projects the pure states of which $\hat{\rho}$ is a statistical mixture (i.e., convex combination) onto $\hat{\Pi}(\hat{A}=\af)(\mathcal{H})$ and assign appropriate probabilities to them. This was demonstrated in detail in Section~\ref{sec:stateUpdate}.

Analogously to quantum systems, a deterministic ontological model represents the proposition $[\hat{A} =\af]$ as the set $\Omega(\hat{A}=\af)$ of ontic states $\lambda$ where this proposition holds, meaning that $\lambda \in \Omega(\hat{A}=\af)$ if and only if $\kappa_{\lambda}[\hat{A}=\af]=1$. Given a state $\mu_{\hat{\rho}}$ and a proposition $[\hat{A} =\af]$, the updated state $\tau_{[\hat{A} = \af]}(\mu)$ modifies $\mu$ to ensure that the proposition $[\hat{A} = \af]$ is true. To achieve this, $\tau_{[\hat{A} = \af]}$ ``projects'' (i.e., conditions) the ontic states of which $\mu$ is a statistical mixture onto $\Omega(\hat{A}=\af)$ and assign appropriate probabilities to them --- this can be made precise with  Choquet's theorem \cite{pedra2023Calgebras}, but this level of precision is unnecessary here. As happens with pure states in quantum mechanics, ontic states where $[\hat{A} = \af]$ holds remain unchanged when updated by $[\hat{A} = \af]$, whereas those in which $[\hat{A} \neq \af]$ is true are lead to the null state (the measure that assigns $0$ to all measurable sets). Note that $[\hat{A} \neq \af]$ is true in a microstate $\lambda$ if and only if $\kappa_{\lambda}[\hat{A}=\af] = 0$. 

We can make this comparison mathematically precise using the total variation distance, which is the canonical notion of distance between probability measures \cite{folland1999real,klenke2014probability}. If $\mu,\nu$ are probability measures in some measurable space $\boldsymbol{\Lambda} \equiv(\Lambda,\mathcal{A})$, the total variation distance between $\mu$ and $\nu$ is the number
\begin{equation}
    \Vert \mu - \nu \Vert \doteq\label{eq:variationDistance} \sup_{\Omega \in \mathcal{A}} \vert \mu(\Omega) - \nu(\Omega) \vert.
\end{equation}
Projecting a state $\psi \in \mathcal{H}$ consists of selecting the element of $\hat{\Pi}(\hat{A} = \alpha)(\mathcal{H})$ that minimises the distance from $\psi$ (Eq.~\eqref{eq:minimizeDistance}). Analogously, conditioning a measure $\mu \in \mathcal{S}(\Lambda)$ on $\Omega[\hat{A} \in \Delta]$ consists of selecting the measure in which $[\hat{A} \in \Delta]$ holds that minimises the distance from $\mu$:

\begin{lemma}\label{lemma:variationDistance}
    Let $\mathcal{O}_{S}$ be a set of observables in some (finite dimensional) quantum system $\mathfrak{S}$, and let $\mathfrak{M} \equiv (\boldsymbol{\Lambda},\Theta,\Psi)$ be a deterministic (state updating) ontological model for $\mathcal{O}_{S}$ (Def.~\ref{def:detOntologicalModel}). Let $\mathcal{S}_{\Lambda}$ be the set of all probability measures on $\boldsymbol{\Lambda}$, and let $\mu \in \mathcal{S}_{\Lambda}$. For any $\hat{A} \in \mathcal{O}_{S}$ and $\Delta \subset \sigma(\hat{A})$, we have
    \begin{equation}
        \Vert \tau_{[\hat{A} \in \Delta]}(\mu) - \mu\Vert =\label{eq:variationDistanceLemma} \min\{\Vert \nu - \mu\Vert: \nu \in \mathcal{S}_{\Lambda}, P_{\nu}[\hat{A} \in \Delta] = 1\},
    \end{equation}
    where $P_{\nu}[\hat{A} \in \Delta] \equiv \nu(f_{\hat{A}}^{-1}(\Delta))$.
\end{lemma}
A proof can be found in the Appendix~\ref{sec:proofs}.

The similarity between the quantum state update and conditional probability is that, by construction, both are optimal methods for reconstructing states to ensure the validity of propositions. The difference --- and this is key --- is that ontic states do not superpose, reducing the update mechanism to a mere enhancement of the experimentalist's knowledge about the system's ontic state. This is because $\Omega(\hat{A} =\af)$ and $\Omega(\hat{A} \neq\af)$ exhaust all ontic states, while the subspaces $\hat{\Pi}(\hat{A}=\af)(\mathcal{H})$ and $\hat{\Pi}(\hat{A}\neq\af)(\mathcal{H})$ does not include all pure states. That is, on the one hand, we have $\Omega(\hat{A} =\af) \cap \Omega(\hat{A} \neq\af) = \emptyset$ and
\begin{equation}
    \Lambda = \Omega(\hat{A} =\af) \cup \Omega(\hat{A} \neq\af).
\end{equation}
On the other hand, we have $\hat{\Pi}(\hat{A}=\af)(\mathcal{H}) \cap \hat{\Pi}(\hat{A}\neq\af)(\mathcal{H}) = \{0\}$ but $\mathcal{H} \neq \hat{\Pi}(\hat{A}=\af)(\mathcal{H}) \cup \hat{\Pi}(\hat{A}\neq\af)(\mathcal{H}) \{0\}$. Instead, $\mathcal{H}$ is the direct sum of these subspaces:
\begin{equation}
    \mathcal{H} = \hat{\Pi}(\hat{A}=\af)(\mathcal{H}) \oplus\hat{\Pi}(\hat{A}\neq\af)(\mathcal{H}).
\end{equation}
As we know, this means that, in some pure states $\psi$, neither $[\hat{A} = \af]$ nor $[\hat{A} \neq \af]$ holds, because these states are superpositions of non-zero vectors $\psi_{\af}$ and $\psi_{\neq \af}$ where $[\hat{A} = \af]$ and $[\hat{A} \neq \af]$ are satisfied, respectively. These are precisely the pure states that undergo a non-trivial update by $[A \in \alpha]$, meaning they are neither left unchanged nor mapped to the zero state. Consequently, any mixed state $\rho = \sum_{i=1}^{m} p_{i} \ketbra{\psi_{i}}$ that contains such a superposition as one of its possible states $\psi_{1},\dots,\psi_{m}$ will undergo a non-trivial update by $[\hat{A} = \af]$, in that it will not simply be conditioned on the set of states $\psi_{i}$ satisfying $P_{\psi_{i}}[\hat{A} = \af]=1$. This is where the classical and quantum update mechanisms diverge, and it is precisely these non-trivial updates  that leaves room for a pivotal aspect of quantum systems: the possibility of order-dependent predictions of sequential measurements, or, equivalently, the existence of incompatible observables. Conditional probability is a fully subjective update, as it only enhances the experimentalist's knowledge about the system's ontic state. The very meaning of Bayes's rule is that, under such subjective/epistemic/statistical update, sequential probabilities are independent of order.

\subsection{On Alternative Updates in Ontological Models}\label{sect:howexistingmodelsupdatethen}

Multiple ontological models have been proposed, which use different rules for state update. Some of them justify their update mechanisms by invoking physical disturbances arising from interactions with measuring apparatuses \cite{harrigan2007interpretation,Hindlycke2022ConOntologicalModel}. For instance, Hindlycke and Larsson's state update rule \cite{Hindlycke2022ConOntologicalModel} effectively combines a basis update (to quote them, an update ensuring ``that the measurement and conjugate contexts remain a symplectic basis'') with a measurement disturbance, where they ``randomise the phase for the (possibly new) $C_k$''. Justifications grounded on measurement disturbances are only tenable if the quantum formalism updates states not through counterfactual propositions but through propositions in the simple past, contradicting our account of the quantum state update rule. Hence, under our account, such ontological models do not properly capture the ontology of quantum systems in the way that we would expect given the nature of the ontological models formalism. It means that, to justify alternative updates in ontological models using measurement disturbances, one must advocate for a reading of the quantum state update mechanism that differs from ours.

Alternatively, some authors propose limits to our capacity to grasp ontic states, thereby restricting the set of measures that can represent quantum states and permitting the violation of conditional probability. This is the case in Spekkens' toy model, for instance \cite{spekkens2007toy}. To understand how such restrictions can affect the state update, suppose that, for some reason (e.g., a ``knowledge balance principle'' dictating that quantum states are measures with ``large'' support, as in the toy model \cite{spekkens2007toy}), there is a limit on which probability measures can represent states in some ontological model $\mathfrak{M} \equiv (\boldsymbol{\Lambda},\Phi,\Psi)$. Let $\mathcal{S}_{Q}(\Lambda) \subsetneq \mathcal{S}(\Lambda)$ be the set of such measures. With this additional constraint, reconstructing a state $\hat{\rho}$ to ensure that a proposition $[\hat{A} \in \Delta]$ holds consists of selecting, within  $\mathcal{S}_{Q}(\Lambda) \subsetneq \mathcal{S}(\Lambda)$, the measure in which $[\hat{A}\in\Delta]$ holds that best approximates $\mu_{\hat{\rho}}$ (for example, one may take the measure in $\mathcal{S}_{Q}(\Lambda) \subsetneq \mathcal{S}(\Lambda)$ satisfying $[\hat{A}\in \Delta]$ that minimises the total variation distance from $\mu_{\hat{\rho}}$).  In summary, the updated state $\tau_{[\hat{A} \in\Delta]}(\mu_{\psi})$ is determined by a constrained optimal approximation, where the constraint arises from the experimentalist's incapacity of grasping the system's ontic state with precision.

The challenge in this, say, ``epistemic'' justification for the violation of conditional probability lies in explaining why sequential probabilities constructed using such a sub-optimal update mechanism would yield well-defined probability measures and correctly predict the relative frequencies observed in sequential measurements. This amounts to assuming that the sub-optimal update mechanism, constrained by the experimentalist's inability to precisely grasp the system's ontic state, correctly predicts the relative frequencies observed when $\hat{\rho}$ is prepared and $\hat{A}$ and $\hat{B}$ are measured in sequence, whereas conditional probability (the unconstrained, optimal epistemological update) would fail. Put differently, an intellect (or ``demon'') free from this epistemological limitation would fail to predict the relative frequencies observed when $\hat{\rho}$ is prepared and $\hat{A}$ and $\hat{B}$ are measured in sequence, implying that the predictive success of quantum theory is rooted in our ignorance. Given we would expect our knowledge of the world not to directly affect the world itself, epistemic constraints seemingly fails to justify the violation of conditional probability in ontological models, at least under our description of the quantum state update. Under alternative descriptions --- allowing, for instance, a change in the system's ontic state --- such justification does remain possible; however, such an account differs from the typical account of the process of state update implicitly assumed in quantum theory.

Another obstacle to alternative updates --- one that does not depend on our particular account, but applies to L\"uders rule as such --- is ensuring that they give rise to conditional probability at the quantum level. As discussed in Section~\ref{sec:compatibility}, the distribution of an observable $\hat{B}$ in quantum mechanics is updated via conditional probability whenever possible, i.e., whenever the proposition $[\hat{A} \in \Delta]$ involves an observable $\hat{A}$ that can be represented as a random variable in the same probability space as $\hat{B}$. This happens when $\hat{A}$ and $\hat{B}$ are compatible. The probability space is given by the joint spectrum of $\hat{A}$ and $\hat{B}$ \cite{landsman2017foundations}, equipped with their joint distribution. Equivalently, it consists of the spectrum of  any observable $\hat{C}$ such that $\hat{A}=f(\hat{C})$ and $\hat{B}=g(\hat{C})$. In this case, the random variables representing $\hat{A}$ and $\hat{B}$ are $f$ and $g$, respectively, and the probability measure is the distribution of $\hat{C}$.  Two questions arise from these well-known features of the quantum formalism: first, how can one construct an update mechanism that, despite not adhering to conditional probability at the ontological level, nonetheless yields conditional probability among compatible observables at the quantum level? This may be conceivable in finite-dimensional systems but is cumbersome in continuous ones. Second, how can we understand the distinction between compatible and incompatible observables, given that ontic states render all observables jointly distributed? Isn't measuring the system's ontic state one way of jointly measuring all observables, just as measuring $\hat{C}$ entails jointly measuring all its functions? 

It may be argued that Beltrametti and Bugajski's model \cite{Beltrametti1995classicalExtension} challenges the line of thought leading to Def.~\ref{def:ontologicalModel}: this is an ontological model that, by construction, updates states in agreement with the quantum state update rule and, for precisely this reason, does not obey conditional probability. In response, we argue that Beltrametti and Bugajski's model is only formally equivalent to Spekkens' ontological models, differing in how measurements are interpreted: if Beltrametti and Bugajski's model obeys the quantum state update rule (as the standard postulates of quantum mechanics demand), then its measurements cannot be interpreted according to the OMF description, namely as leading ``to an update in one’s information about what the ontic state of the system was prior to the procedure being implemented'' \cite{spekkens2005contextuality}.

In fact, in Beltrametti and Bugajski's model, the set of ontic states consists of all pure quantum states of the system. Each convex decomposition of a state $\hat{\rho}$ defines a probability measure representing it in the model (these measures are not necessarily equal, meaning that the model is generalised contextual for preparations \cite{spekkens2005contextuality}). The Markov Kernel associated with an observable $\hat{A}$ assigns probabilities to pure states via the Born rule, that is to say, $\kappa_{\psi}[\hat{A}=\af] \doteq \langle \psi | \hat{\Pi}(\hat{A}=\af) \vert \psi \rangle$. Finally, the model recovers the quantum predictions via Eq.~\eqref{eq:probabilityOntologicalModelPre}, which is equivalent to the Born rule. Interpreting measurements as updating one’s information about what was the ontic/pure state prior to measurement in this model is equivalent to assuming that such interpretation holds in quantum mechanics, which is recognised as an incorrect view --- it conflicts, to mention only one issue, with state superposition. Therefore, this model does not invalidate our claim that ontological models, as conceived in the generalised contextuality literature \cite{spekkens2005contextuality}, must update states via conditional probability.

It is important to recall that Beltrametti and Bugajski's work predates the development of generalised contextuality; their reformulation of quantum mechanics was intended to demonstrate that it admits a certain classical extension (defined in a mathematically precise sense), a goal they successfully achieved. Moreover, Spekkens was the first to point out that their model does not qualify as a classical one according to the OMF criterion, as it is contextual for preparations \cite{spekkens2005contextuality}.

To conclude, we would like to emphasise that, in continuous systems, Eq.~\eqref{eq:selfConsistencyDist} is virtually sufficient to rule out any update that differs from conditional probability in deterministic models --- if you accept our analysis of state update. As discussed in the end of Section~\ref{sec:analyzingModels}, this is because measurements in ontological models are  thought of as ``measurements \textit{of} the ontic state of the system'' which ``might only enable one to infer probabilities for the system to have been in different ontic states'' \cite{spekkens2005contextuality}, suggesting that there exists a (possibly multi-valued) observable $\hat{A}$ that simply reveals in which ontic state the system exists. Every measurable set  $\Omega \subset \Lambda$ satisfies $\Omega=f_{\hat{A}}^{-1}(\Delta)$ for some $\Delta$ (just get $\Delta \doteq f_{\hat{A}}(\Omega)$), thus, given any measurable set $\Omega$, conditioning a state $\mu_{\hat{\rho}}$ on $\Omega$ consists of updating the state $\hat{\rho}$ with a proposition $[\hat{A} \in \Delta]$. 

\subsection{Concluding Remarks}\label{sec:concludingRemarks}

In this paper, we have shown that ontological models (crafted as per Spekkens' ontological models framework), which admit sequential measurement of incompatible observables, cannot represent the state update we would expect given standard conditioning of probabilities. Inversely, ontological models which can reproduce the stated update we would expect from conditional probabilities can only represent sequential measurement of pairwise compatible observables. We argued that ontological models consistent with the canonical postulates of quantum mechanics must update states via conditional probability, from which it follows that incompatibility --- or equivalently, order-dependent predictions for sequential measurements --- is sufficient to rule out ontological models of quantum systems.

It is natural to ask what impact our result has on OMF-based definitions, such as generalized contextuality and recent approaches to the ontology problem \cite{leifer2014real}. A first question is whether these categories remain applicable to quantum mechanics for those who accept the counterfactual account we defend. Another promising direction is to identify examples in classical physics of theories with incompatible observables that employ such a counterfactual update and are therefore impossible to represent within ontological models. Such examples would directly challenge the goal of making generalized contextuality a notion of classicality that is universally applicable \cite{schmid2021characterization, schmid2024structureTheorem, schmid2024inequalities, schmid2024objections}, making this an important avenue of research. We aim to explore these directions in future work. In addition, we plan to extend our result to continuous and infinite-dimensional systems, investigate its intersection with hidden-variable models constructed using negative (quasi)probabilities, and examine its implications for the notion of theory-independent nonlocality \cite{brunner2014bell,popescu2014beyond}.

Future work will aim to prove our result for continuous and infinite-dimensional systems; look at the intersection of our result with hidden variable models constructed with negative (quasi)probabilities for systems either being in a certain ontic state $\lambda$ given they have a certain quantum state (i.e., negative $\mu$s), or a system returning a value in a given range $\Delta$ under measurement of a given observable $\hat{A}$ when in a given ontic state $\lambda$ (i.e., negative $\kappa$s); and explore the implications of our results for the notion of theory-independent nonlocality \cite{brunner2014bell,popescu2014beyond}.

\begin{acknowledgments}
We would like to thank Jan-\AA ke Larsson, Rafael Wagner, and Arthur Fine for useful discussions, and comments on an early draft of this paper. AT acknowledges support from the National Council for Scientific and Technological Development (CNPq) (Grant Number GD 142295/2020-5), and the project ``Implementação do sistema CV-QKD em uma rede quântica real'', supported by Quantum Industrial Innovation (QuIIN) — EMBRAPII CIMATEC Competence Center in Quantum Technologies, with financial resources from the PPI IoT/Manufatura 4.0 of the MCTI, grant number 053/2023, signed with EMBRAPII. BA acknowledges support from São Paulo Research Foundation (FAPESP) and Serrapilheira Institute. JRH acknowledges support from a Royal Society Research Grant (RG/R1/251590), and from their EPSRC Quantum Technologies Career Acceleration Fellowship (UKRI1217).
\end{acknowledgments}

\bibliography{Bibliography}

\appendix

\section{Mathematical Preliminaries}\label{sec:measureTheory}

In this section, we summarise the measure-theoretic results used throughout the manuscript. Detailed presentations can be found in Refs.~\cite{halmos1974measure,tao2011measure, cohn2013measure,klenke2014probability} and elsewhere.

In probability theory, measurable sets represent events to which probabilities can be assigned. The complement $\Lambda \backslash \Omega$ of an event $\Omega$ corresponds to its negation, i.e., the event that $\Omega$ does not occur. The union of events, on the other hand, corresponds to the event in which some of them occur. The set $\Lambda$ is said to be the sample space, and its elements are called sample points.

Probability spaces formalise the concept of probability for continuous variables. Let $\Lambda$ be a non-empty set, and let $\mathcal{A}$ be a collection of subsets of $\Lambda$. Then $\mathcal{A}$ is said to be a \textit{$\sigma$-algebra} on $\Lambda$ if it satisfies the following conditions.
\begin{itemize}
    \item[(i)] Closure under complementations: if $\Omega \in \mathcal{A}$ then $\Lambda \backslash \Omega \in \mathcal{A}$.
    \item[(ii)] Closure under countable unions: For any sequence $\Omega_{i} \in \mathcal{A}$, $i \in \mathbb{N}$,
    \begin{equation}
        \bigcup_{i=1}^{\infty} \Omega_{i} \in \mathcal{A}.
    \end{equation}
    \item[(iii)] $\Lambda \in \mathcal{A}$.
\end{itemize}
Note that items (a) and (c) entails that $\emptyset \in \mathcal{A}$, which in turn, together with item (b), ensures that $\mathcal{A}$ is closed under finite unions. It is also easy to show that $\sigma$-algebras are closed under countable intersections \cite{tao2011measure, halmos1974measure}.

If $\mathcal{A}$ is a $\sigma$-algebra on $\Lambda$, the pair $\boldsymbol{\Lambda} \equiv (\Lambda,\mathcal{A})$ is said to be a \textit{measurable space}. The elements of $\mathcal{A}$ are called the measurable sets of $\boldsymbol{\Lambda}$. 

Probabilities are assigned to events via probability measures. These are functions $\mathcal{A} \ri [0,1]$ that respect the $\sigma$-algebra structure. The definition of measure is as follows.

A (positive) \textit{measure} on a measurable space $\boldsymbol{\Lambda} \equiv (\Lambda,\mathcal{A})$ is a function $\mu:\mathcal{A} \ri [0,\infty]$ satisfying the following conditions.
\begin{itemize}
    \item[(a)] $\sigma$-additivity: if $\Omega_{i} \in \mathcal{A}$, $i \in \mathbb{N}$, are pairwise disjoint, then
    \begin{equation}
        \mu(\bigcup_{i=1}^{\infty} \Omega_{i}) =\label{eq:sigmaAdditivity} \sum_{i=1}^{\infty} \mu(\Omega_{i}).
    \end{equation}
    \item[(b)] $\mu(\emptyset) = 0$
\end{itemize}
If $\mu$ is normalised, i.e., if $\mu(\Lambda)=1$, $\mu$ is said to be a \textit{probability measure} on $\boldsymbol{\Lambda}$. The number $\mu(\Omega)$ is called the measure of the measurable set $\Omega$. In the particular case where $\mu$ is a probability measure, $\mu(\Omega)$ is known as the probability of $\Omega$.

A \textit{measure space} consists of a measurable space equipped with a measure. That is, a measure space is a triple $(\Lambda,\mathcal{A},\mu)$, where $\boldsymbol{\Lambda} \equiv (\Lambda,\mathcal{A})$ is a measurable space and $\mu$ is a measure on $\boldsymbol{\Lambda}$. A \textit{probability space} is a measure space whose measure is normalised (i.e., a probability measure). 

It is easy to show that measures are monotonic and subadditive. Monotonicity means that, if two measurable sets $\Omega_{1},\Omega_{2}$ satisfy $\Omega_{1} \subset \Omega_{2}$, then $\mu(\Omega_{1}) \leq \mu(\Omega_{2})$. Consequently, every event in a probability space has measure (i.e., probability) between $0$ and $1$, as expected. Subadditivity means that, for any sequence of sets $\Omega_{i} \in \mathcal{A}$, $i \in \mathbb{N}$, we have
\begin{equation}
    \mu(\bigcup_{i=1}^{\infty} \Omega_{i}) \leq\label{eq:subadditivity} \sum_{i=1}^{\infty} \mu(\Omega_{i}). 
\end{equation}

In probability theory, Eq.~\eqref{eq:sigmaAdditivity} ensures that, if $\Omega_{i} \in \mathcal{A}$, $i \in \mathcal{N}$, are mutually exclusive events, the probability that some of them occur is just the sum of their probabilities. If the events are not mutually exclusive, the sum of their probabilities may be larger than the probability that some of them occur (due to the potential existence of intersecting events), as encapsulated in Eq.~\eqref{eq:subadditivity}.


A central topic in measure theory is integration theory, particularly Lebesgue integration theory \cite{halmos1974measure, tao2011measure, klenke2014probability}. It is out of the scope of this work to provide a detailed presentation of Lebesgue's definition of integral --- such discussions can be found in any undergraduate textbook on measure theory (see e.g. Refs.~\cite{tao2011measure, halmos1974measure}) --- but the core idea behind it is easy to grasp. The construction begins with simple functions, which are linear combinations of indicator functions of measurable sets. More precisely, a function $f:\Lambda \ri \R$ is said to be simple in the measurable space $(\Lambda,\mathcal{A})$ if there are pairwise disjoint measurable sets $\Omega_{1},\dots,\Omega_{m}$ and real numbers $\beta_{1},\dots,\beta_{m}$ such that $f=\sum_{i=1}^{m} \beta_{i}\chi_{\Omega_{i}}$, where $\chi_{\Omega_{i}}$ is the indicator function of $\Omega_{i}$. Now let $\mu$ be a measure in $\boldsymbol{\Lambda} \equiv (\Lambda,\mathcal{A})$. The Lebesgue integral of $f$ with respect to $\mu$ (equivalently, in the measure space $(\Lambda,\mathcal{A},\mu)$) is the number
\begin{equation}
    \int_{\Lambda} f(\lambda) \ \mu(d\lambda) \doteq \sum_{i=1}^{m} \beta_{i} \mu(\Omega_{i}).
\end{equation}
The integral of more complicated real-valued measurable functions is defined through converging sequences of simple functions. Roughly speaking, it consists in partitioning the codomain of the function into subintervals and taking the limit where their length approaches zero, similarly to what the Riemann integral does to the function's domain \cite{folland1999real}. The Lebesgue integral of a complex-valued function is defined via their real and imaginary part. 


The Lebesgue integral generalises the proper Riemann integral in its domain of applicability. Every proper Riemann integral on an interval $[\af,\beta] \subset \R$ is a Lebesgue integral w.r.t. the so-called Lebesgue measure (the measure $\mu_{\mathfrak{L}}$ on $\R$ that assign lengths to intervals, i.e., $\mu_{\mathfrak{L}}([\af,\beta]) \doteq \beta - \af$ for each interval $[\af,\beta] \subset \R$). In addition, some functions $f:[\af,\beta] \ri \R$ which are not Riemann integrable have a well-defined Lebesgue integral. This is because a bounded function is Riemann integrable on $[\af,\beta]$ if and only if it is continuous almost-everywhere (the set of points where it is discontinuous has Lebesgue measure equals $0$), a condition that does not apply to the Lebesgue integral. The canonical example is the indicator function of the rational numbers, which is discontinuous in every point of $[0,1]$ but has a well-defined Lebesgue integral, which is $0$.

To conclude, let's state the Kolmogorov extension theorem, which is central for our proof of Proposition~\ref{prop:impossibility}. For simplicity, we will restrict it to probability spaces with finite sample spaces. The general version can be found in Ref. \cite{tao2011measure} and elsewhere.

\begin{theorem}[Kolmogorov extension theorem for finite spaces]\label{thm:kolmogorov} Let $\Lambda_{A}$, $A \in \mathcal{O}$, be finite sets, where $\mathcal{O}$ is a nonempty set of indexes. Let $\Lambda$ be the Cartesian product $\prod_{A\in \mathcal{O}} \Lambda_{A}$, namely the collection of all tuples $\lambda \equiv (\lambda_{A})_{A \in \mathcal{O} }$ satisfying $\lambda_{A} \in \Lambda_{A}$ for each $A \in \mathcal{O}$.
For each finite sequence $A_{1},\dots,A_{m}\in \mathcal{O}$, let $P[(A_{1},\dots,A_{m}) \in \cdot \ ]$ be a probability measure on the power set of $\prod_{i=1}^{m}\Lambda_{A_{i}}$. Suppose that, for any finite sequence $A_{1},\dots,A_{m} \in \mathcal{O}$, the following conditions are satisfied.
\begin{itemize}
    \item[(a)] For any permutation $\pi$ of $\{1,\dots,m\}$ and any sets $\Delta_{i} \subset \Lambda_{A_{i}}$,
    \begin{equation}
        P[A_{1} \in \Delta_{1},\dots,A_{1} \in \Delta_{m}] = P[A_{\pi(1)} \in \Delta_{\pi(1)},\dots,A_{\pi(1)} \in \Delta_{\pi(m)}],
    \end{equation}
    where $P[A_{1} \in \Delta_{1},\dots,A_{1} \in \Delta_{m}] \equiv P[(A_{1},\dots,A_{m}) \in \Delta_{1}\times\dots\times\Delta_{m}]$, and analogously for the right hand side of the equation.
    \item[(b)] For any subsequence $A_{i_{1}},\dots,A_{i_{n}}$ of $A_{1},\dots,A_{m}$,  the measure $P[(A_{i_{1}},\dots,A_{i_{n}}) \in \cdot \ ]$ corresponds to the measure $P[(A_{1},\dots,A_{m}) \in \cdot \ ]$ marginalised over all indexes in $\{\hat{A}_{1},\dots,\hat{A}_{m}\}$ except $\hat{A}_{n_{1}},\dots,\hat{A}_{n_{M}}$. That is to say, for any sets $\Delta_{i_{j}} \subset \Lambda_{A_{i_{j}}}$, $j=1,\dots,n$, we have
    \begin{equation}
        P[A_{i_{1}} \in \Delta_{i_{i}},\dots,A_{i_{n}} \in \Delta_{i_{n}}] = P[A_{1} \in \Delta'_{1},\dots A_{m} \in \Delta'_{m}],
    \end{equation}
    where
    \begin{equation}
        \Delta'_{i} = \begin{cases}
            \Delta_{i_{j}} \ \text{if} \ i=i_{j} \text{for some} \ j\\
            \Lambda_{A_{i}} \ \text{otherwise}.
        \end{cases}
    \end{equation}
\end{itemize}
Then there exists a probability measure $P$ on the product $\sigma$-algebra \cite{tao2011measure} on $\Lambda \equiv \prod_{A\in \mathcal{O}} \Lambda_{A}$ such that, for any finite sequence  $A_{1},\dots,A_{m} \in \mathcal{O}$ and sets $\Delta_{i} \subset \Lambda_{A_{i}}$, $i=1,\dots,m$, we have
\begin{equation}
    P[A_{1} \in \Delta_{1},\dots,A_{m}\in\Delta_{m}] =\label{eq:kolmogorovViaSet} P[\{\lambda \in \Lambda: \forall_{i}(\lambda_{i} \in \Delta_{i})\}]. 
\end{equation}
\end{theorem}

To compare Eq.~\eqref{eq:kolmogorovViaSet} with the Proof of Proposition~\ref{prop:impossibility}, note that 
\begin{equation}
    \{\lambda \in \Lambda: \forall_{i}(\lambda_{i} \in \Delta_{i})\} = \bigcap_{i=1}^{m}f_{A_{i}}^{-1}(\Delta_{i}),
\end{equation}
where $f_{A_{i}}:\Lambda \ri \Lambda_{A_{i}}$ is the coordinate projection function $\lambda \mapsto \lambda_{A_{i}}$.

\section{Proofs}\label{sec:proofs}

\noindent \textbf{Proof of Lemma~\ref{lemma:compatibility}:}
Let $\hat{A}$ and $\hat{B}$ be (not necessarily compatible) observables in a finite-dimensional quantum system, and let $\hat{A}=\sum_{\af \in \sigma(\hat{A})}\af \hat{\Pi}(\hat{A}=\alpha)$ and $\hat{B}=\sum_{\beta \in \sigma(\hat{B})}\beta \hat{\Pi}(\hat{B} = \beta)$ be their spectral decompositions \cite{landsman2017foundations}. For any state $\hat{\rho}$ and  sets $\Delta \subset \sigma(\hat{A})$, $\Sigma \subset \sigma(\hat{B})$, define
\begin{align}
    P_{\hat{\rho}}[\hat{A} \in \Delta,\hat{B} \in \Sigma] &\equiv P_{\hat{\rho}}[(\hat{A},\hat{B}) \in \Delta \times \Sigma]
    \\
    &\doteq P_{\hat{\rho}}[\hat{A} \in \Delta]P_{\hat{\rho}}[\hat{B} \in \Sigma | \hat{A} \in \Delta]. 
\end{align}
As discussed in Section~\ref{sec:compatibility}, it induces a probability measure $P_{\hat{\rho}}[(\hat{A},\hat{B}) \in  \cdot \ ]$ on $\R^{2}$. By construction, the support of this measure is a subset of the finite set $\sigma(\hat{A}) \times \sigma(\hat{B})$. Denote by $P_{\hat{\rho}}[(\hat{A},\hat{B}) = \cdot \ ]$ the probability distribution on $\sigma(\hat{A}) \times \sigma(\hat{B})$ associated with $P_{\hat{\rho}}[(\hat{A},\hat{B}) \in  \cdot \ ]$, i.e., $P_{\hat{\rho}}[\hat{A}=\af,\hat{B}=\beta] \doteq P_{\hat{\rho}}(\hat{A} \in \{\af\},\hat{B} \in \{\beta\})$. As shown in Eq.~\eqref{eq:sequentialDistributionQuantum},
\begin{equation}
    P_{\hat{\rho}}[\hat{A}=\af,\hat{B}=\beta] = \Exp{\hat{\Pi}(\hat{A}=\alpha)\hat{\Pi}(\hat{B} = \beta)\hat{\Pi}(\hat{A}=\alpha)}
\end{equation}
States separate observables in quantum systems, i.e., two self-adjoint operators $C,D$ are equal if and only if $\Exp{C}=\Exp{D}$ for each state $\hat{\rho}$ \cite{pedra2023Calgebras}. Therefore, the probability distributions $P_{\hat{\rho}}[(\hat{A},\hat{B}) =  \cdot \ ]$ and $P_{\hat{\rho}}[(\hat{B},\hat{A}) =  \cdot \ ]$  are equal (up to a permutation) for every state $\hat{\rho}$ if and only if $\hat{\Pi}(\hat{A}=\alpha)\hat{\Pi}(\hat{B} = \beta)\hat{\Pi}(\hat{A}=\alpha)= \hat{\Pi}(\hat{B} = \beta)\hat{\Pi}(\hat{A}=\alpha)\hat{\Pi}(\hat{B} = \beta)$ for any $(\af,\beta) \in \sigma(\hat{A}) \times \sigma(\hat{B})$. As shown in Ref.~\cite{rehder1980commute}, this is equivalent to saying that $\hat{\Pi}(\hat{A}=\alpha)$ and $\hat{\Pi}(\hat{B} = \beta)$ are compatible. We know that $\hat{A}$ and $\hat{B}$ are compatible if and only if $\hat{\Pi}(\hat{A}=\alpha)$ and $\hat{\Pi}(\hat{B} = \beta)$ are compatible for each $(\af,\beta) \in \sigma(\hat{A}) \times \sigma(\hat{B})$ \cite{nielsen2000quantum}. Hence, $\hat{A}$ and $\hat{B}$ are compatible if and only if, for each state $\hat{\rho}$ and each $(\af,\beta) \in \sigma(\hat{A}) \times \sigma(\hat{B})$,
\begin{equation}
    P_{\hat{\rho}}[\hat{A}=\af,\hat{B}=\beta]=P_{\hat{\rho}}[\hat{B}=\beta,\hat{A} =\af].
\end{equation}
This is equivalent to saying that, for every state $\hat{\rho}$, the probability measures $P_{\hat{\rho}}[(\hat{A},\hat{B}) \in  \cdot \ ]$ and $P_{\hat{\rho}}[(\hat{A},\hat{B}) \in  \cdot \ ]$ are equal up to a permutation, which in turn is equivalent to saying that Eq.~\eqref{eq:bayesLemma} holds for any sets $\Delta \subset \sigma(\hat{A})$ and $\Sigma \subset \sigma(\hat{B})$.

\hfill\ $\square \medskip$

\noindent \textbf{Proof of Lemma~\ref{lemma:ontologicalModel}:}
Let $\mathcal{O}_{S}$ be a non-empty set of observables in a finite-dimensional quantum system $\mathfrak{S}$, and let $\mathfrak{M} \equiv (\boldsymbol{\Lambda},\Phi,\Psi)$ be a state-updating ontological model for $\mathcal{O}_{S}$. Let $\hat{A} \in \mathcal{O}_{S}$, and let $\hat{A}=\sum_{\af \in \sigma(\hat{\hat{A}})} \af \hat{\Pi}(\hat{A}=\alpha)$ be its spectral decomposition. Recall that, for any function $g:\sigma(\hat{A})\ri \R$, we have $g(\hat{A}) = \sum_{\af \in \sigma(\hat{A})} g(\af) \hat{\Pi}(\hat{A}=\alpha)$. It thus follows from the linearity of the Lebesgue integral and from the definition of integral of simple functions that
\begin{equation} \begin{split}
    \Exp{g(\hat{A})} &= \sum_{\af \in \sigma(\hat{A})} g(\af) \Exp{\hat{\Pi}(\hat{A}=\alpha)}
    \\
    &=\sum_{\af \in \sigma(\hat{A})} g(\af) P_{\hat{\rho}}[\hat{A} = \af] 
    \\
    &= \sum_{\af \in \sigma(\hat{A})} g(\af) \int_{\Lambda} \kappa_{\lambda}[\hat{A}=\af] \ \mu_{\hat{\rho}}(d\lambda)
    \\
    &= \int_{\Lambda} \left(\sum_{\af \in \sigma(\hat{A})} g(\af) \kappa_{\lambda}[\hat{A} = \af]\right) \ \mu_{\hat{\rho}}(d\lambda)
    \\
    &= \int_{\Lambda}\int_{\sigma(\hat{A})}g(\hat{A}) \ d\kappa_{\lambda}[\hat{A} \in d\af]\ \mu_{\hat{\rho}}(d\lambda)
\end{split}\end{equation}
This proves Eq.~\eqref{eq:expectationOfFunctionModel}. Eq.~\eqref{eq:expectationModel} follows from the fact that $\hat{A}=\text{id}(\hat{A})$, where $\text{id}$ is the identity function $\af \mapsto \af$. 

Next, let $A_{1},\dots,A_{m}$ be pairwise compatible observables and $\Delta_{i} \subset \sigma(\hat{A}_{i})$, $i=1,\dots,m$, be sets. For simplicity, write $P_{\hat{\rho}}[\hat{A}_{i} \in \Delta_{i}| \dots,\hat{A}_{i-1} \in \Delta_{i-1}]$ rather than $P_{\hat{\rho}}[\hat{A}_{i} \in \Delta_{i}|\hat{A}_{1} \in \Delta_{1}, \dots,\hat{A}_{i-1} \in \Delta_{i-1}]$ (other shorthands are hopefully self explanatory). It follows from Eqs.~\eqref{eq:finiteSequentialMeasure}, \eqref{eq:standardJoint} and \eqref{eq:conditionalProbabilityModel} that
\begin{equation}
\begin{split}
    P_{\hat{\rho}}[\hat{A}_{1} \in \Delta_{1},\dots,\hat{A}_{m} \in \Delta_{m}] &=\prod_{i=1}^{m-1} P_{\hat{\rho}}[\hat{A}_{i} \in \Delta_{i}| \dots,\hat{A}_{i-1} \in \Delta_{i-1}]\\
    &\times\int_{\Lambda} \kappa_{\lambda}[\hat{A}_{m} \in \Delta_{m}] \mu_{\hat{\rho}}(d\lambda | \dots,\hat{A}_{m-1} \in \Delta_{m-1}]\\
    &= \prod_{i=1}^{m-1} P_{\hat{\rho}}[\hat{A}_{i} \in \Delta_{i}| \hat{A}_{1} \in \Delta_{1},\dots,\hat{A}_{i-1} \in \Delta_{i-1}]\\
    &\times\int_{\Lambda} \frac{\kappa_{\lambda}[\hat{A}_{m} \in \Delta_{m}]\kappa_{\lambda}[\hat{A}_{m-1} \in \Delta_{m-1}]}{P_{\hat{\rho}}[\hat{A}_{m-1} \in \Delta_{m-1}|\dots,\hat{A}_{m-2} \in \Delta_{m-2}]} \mu_{\hat{\rho}}(d\lambda |\dots,\hat{A}_{m-1} \in \Delta_{m-2}]
    \\
    &= \prod_{i=1}^{m-2} P_{\hat{\rho}}[\hat{A}_{i} \in \Delta_{i}| \hat{A}_{1} \in \Delta_{1},\dots,\hat{A}_{i-1} \in \Delta_{i-1}]\\
    &\times\int_{\Lambda} \kappa_{\lambda}[\hat{A}_{m} \in \Delta_{m}]\kappa_{\lambda}[\hat{A}_{m-1} \in \Delta_{m-1}]\mu_{\hat{\rho}}(d\lambda |\dots,\hat{A}_{m-2} \in \Delta_{m-2}]
    \\
    &= \prod_{i=1}^{m-3} P_{\hat{\rho}}[\hat{A}_{i} \in \Delta_{i}| \hat{A}_{1} \in \Delta_{1},\dots,\hat{A}_{i-1} \in \Delta_{i-1}]\\
    &\times\int_{\Lambda} \prod_{i=m-2}^{m}\kappa_{\lambda}[\hat{A}_{i} \in \Delta_{i}]\mu_{\hat{\rho}}(d\lambda |\dots,\hat{A}_{m-3} \in \Delta_{m-3}]
    \\
    &= \dots
    \\
    &= \int_{\Lambda} \prod_{i=1}^{m}\kappa_{\lambda}[\hat{A}_{i} \in \Delta_{i}]\mu_{\hat{\rho}}(d\lambda).
\end{split}
\end{equation}
(recall that $\int g d\left(\int f d\mu\right) = \int fg d\mu$). It proves item (b). To conclude, let $\hat{A}$ and $\hat{B}$ be compatible observables, and let $g: \R^{2} \ri \R$ be a  function. Recall that the observable $g(\hat{A},\hat{B})$ is given by
\begin{equation}
    g(\hat{A},\hat{B}) = \sum_{\af \in \sigma(\hat{A})} \sum_{\beta \in \sigma(\hat{B})}g(\af,\beta) \hat{\Pi}(\hat{A}=\alpha) \hat{\Pi}(\hat{B} = \beta),
\end{equation}
where  $\hat{A}=\sum_{\af \in \sigma(\hat{A})}\af \hat{\Pi}(\hat{A}=\alpha)$ and $\hat{B}=\sum_{\beta \in \sigma(\hat{B})} \beta \hat{\Pi}(\hat{B} = \beta)$ are the spectral decompositions of $\hat{A}$ and $\hat{B}$. For each state $\hat{\rho}$,
\begin{equation} \begin{split}
    \Exp{g(\hat{A},\hat{B})}&= \sum_{\af,\beta} g(\af,\beta) \Exp{\hat{\Pi}(\hat{A}=\alpha)\hat{\Pi}(\hat{B} = \beta)}
    \\
    &=\sum_{\af,\beta} g(\af,\beta) P_{\hat{\rho}}[\hat{A}=\af,\hat{B}=\beta]
    \\
    &=\label{eq:expJointFunctionIntegral} \sum_{\af,\beta} g(\af,\beta)\int_{\Lambda} \kappa_{\lambda}[\hat{A}=\af]\kappa_{\lambda}[\hat{B}=\beta] \ \mu_{\hat{\rho}}(d\lambda)
    \\
    &= \sum_{\af} \int_{\Lambda} \int_{\sigma(\hat{B})}g(\af,\beta)\kappa_{\lambda}[\hat{A}=\af]\ \kappa_{\lambda}[\hat{B} \in d\beta] \mu_{\hat{\rho}}(d\lambda)
    \\
    &= \int_{\Lambda} \int_{\sigma(\hat{A})}\int_{\sigma(\hat{B})}g(\af,\beta)\kappa_{\lambda}[\hat{A} \in d\af]\ \kappa_{\lambda}[\hat{B} \in d\beta] \mu_{\hat{\rho}}(d\lambda)
\end{split}\end{equation}
Since $\hat{A}+\hat{B} = g_{+}(\hat{A},\hat{B})$ and $\hat{A}\hat{B} = g_{\circ}(\hat{A},\hat{B})$, with $g_{+}(\af,\beta) \doteq \af +\beta$ and $g_{\circ}(\af,\beta) \doteq \af \beta$, item (c) follows.  
\hfill\ $\square \medskip$

\noindent \textbf{Proof of Corollary~\ref{cor:DeterministicModel}:} Let $\mathfrak{S}$ be a finite-dimensional quantum system, and let $\mathcal{O}_{S}$ be a non-empty set of observables of $\mathfrak{S}$. Let $\mathfrak{M} \equiv (\boldsymbol{\Lambda},\Phi,\Psi)$ be a deterministic ontological model for $\mathcal{O}_{S}$. For each observable $\hat{B} \in \mathcal{O}_{S}$ and set $\Sigma \subset \sigma(\hat{B})$, let $\kappa_{( \cdot )}[\hat{B} \in \Sigma]$ be the indicator function of $f_{\hat{B}}^{-1}(\Sigma)$, i.e., $\kappa_{\lambda}[\hat{B} \in \Sigma] = 1$ if $f_{\hat{B}}(\lambda) \in \Sigma$ and $\kappa_{\lambda}[\hat{B} \in \Sigma] = 0$ otherwise. Similarly, let $\kappa_{(\cdot )}[\hat{B}=\beta]$ be the indicator function of $f_{\hat{B}}^{-1}(\{\beta\})$, where $\beta \in \sigma(\hat{B})$. Recall that $\kappa_{(\cdot )}[\hat{B} \in \cdot \ ]$ is the Markov kernel determined by the random variable $f_{\hat{B}}:\Lambda \ri \R$. By construction, we have
\begin{equation}
    \begin{split}
        f_{\hat{B}}(\lambda) &= \sum_{\beta \in \sigma(\hat{B})} \beta \kappa_{\lambda}[\hat{B}=\beta]
        \\
        &=\label{eq:deterministicViaKernel} \int_{\sigma(\hat{B})} \beta \kappa_{\lambda}[\hat{B} \in d\beta].
    \end{split}
\end{equation}

Item (a) of Corollary~\ref{cor:DeterministicModel} immediately follows from Eqs.~\eqref{eq:expectationModel} and \eqref{eq:deterministicViaKernel}. Item (b) follows from the trivial fact that the product of indicator functions is the indicator function of intersection of their support. That is, for any observables $\hat{A}_{1},\dots,\hat{A}_{m} \in \mathcal{O}_{S}$ and sets $\Delta_{i} \subset \sigma(\hat{A})$, we have
\begin{equation}
    \prod_{i=1}^{m}\kappa_{\lambda}[\hat{A}_{i} \in \Delta_{i}] = \chi_{\bigcap_{i=1}^{m}f_{\hat{A}}^{-1}(\Delta_{i})}(\lambda),
\end{equation}
where $\chi_{\bigcap_{i=1}^{m}f_{\hat{A}}^{-1}(\Delta_{i})}$ denotes the indicator function of $\bigcap_{i=1}^{m}f_{\hat{A}}^{-1}(\Delta_{i})$. It follows from the definition of Lebesgue integral that the integral of $\chi_{\bigcap_{i=1}^{m} f_{\hat{A}}^{-1}(\Delta_{i})}$ w.r.t. $\mu_{\hat{\rho}}$ is $\mu_{\hat{\rho}}(\bigcap_{i=1}^{m}f_{\hat{A}}^{-1}(\Delta_{i}))$, thus item (b) holds. Finally, let $\hat{A},\hat{B} \in \mathcal{O}_{S}$ be compatible observables, and let $g: \sigma(\hat{A}) \times \sigma(\hat{B}) \ri \R$ be a  function. It is easy to see that
\begin{equation}
    \begin{split}
        (g \circ (f,g))(\lambda) &= \sum_{\af \in \sigma(\hat{A})}\sum_{\beta \in \sigma(\hat{B})} g(\af,\beta)\kappa_{\lambda}[\hat{A}=\af]\kappa_{\lambda}[\hat{B}=\beta], 
    \end{split}
\end{equation}
where $(f,g)(\lambda) \doteq (f(\lambda),g(\lambda))$ for each $\lambda \in \Lambda$. Hence, it follows from Eq.~\eqref{eq:expJointFunctionIntegral} that
\begin{equation} \begin{split}
    \Exp{g(\hat{A},\hat{B})}&=  \sum_{\af,\beta} g(\af,\beta)\int_{\Lambda} \kappa_{\lambda}[\hat{A}=\af]\kappa_{\lambda}[\hat{B}=\beta] \ \mu_{\hat{\rho}}(d\lambda)
    \\
    &=  \int_{\Lambda} \sum_{\af,\beta} g(\af,\beta)\kappa_{\lambda}[\hat{A}=\af]\kappa_{\lambda}[\hat{B}=\beta] \ \mu_{\hat{\rho}}(d\lambda)
    \\
    &=  \int_{\Lambda} (h \circ (f,g))(\lambda) \ \mu_{\hat{\rho}}(d\lambda)
\end{split}\end{equation}
Since $\hat{A}+\hat{B} = g_{+}(\hat{A},\hat{B})$ and $\hat{A}\hat{B} = g_{\circ}(\hat{A},\hat{B})$, with $g_{+}(\af,\beta) \doteq \af +\beta$ and $g_{\circ}(\af,\beta) \doteq \af \beta$, and since $g_{+} \circ (f_{\hat{A}},f_{\hat{B}}) = f_{\hat{A}} + f_{\hat{B}}$ and  $g_{\circ} \circ (f_{\hat{A}},f_{\hat{B}}) = f_{\hat{A}} \cdot f_{\hat{B}}$, item (c) follows.  
\hfill\ $\square \medskip$

\noindent \textbf{Proof of Proposition~\ref{prop:impossibility}:} 
    Let $\mathcal{O}_{S}$ be a set of observables in a finite-dimensional quantum system $\mathfrak{S}$. If an ontological model for $\mathcal{O}_{S}$ exists, Lemma~\ref{lemma:compatibility} and item (b) of Lemma~\ref{lemma:ontologicalModel} entail that the observables of $\mathcal{O}_{S}$ are pairwise compatible. Therefore, item (a) implies item (c). Item (a) follows from item (b) by definition. Hence, all that remains for us to prove is that (b) follows from (c). 
    
    Suppose that $\mathcal{O}_{S}$ is a set of pairwise compatible observables. Let $\Lambda$ be the Cartesian product $\prod_{\hat{A} \in \mathcal{O}_{S}} \sigma(\hat{A})$, and let $\mathcal{A}$ be the product $\sigma$-algebra $\prod_{\hat{A} \in \mathcal{O}_{S}}\mathfrak{P}(\sigma(\hat{A}))$ \cite{tao2011measure}, where $\mathfrak{P}(\sigma(\hat{A}))$ denotes the collection of all subsets of $\sigma(\hat{A})$. Recall that elements of $\Lambda \equiv \prod_{\hat{A} \in \mathcal{O}_{S}} \sigma(\hat{A})$ are tuples $\lambda \equiv (\lambda_{\hat{A}})_{\hat{A} \in \mathcal{O}_{S} }$ satisfying $\lambda_{\hat{A}} \in \sigma(\hat{A})$ for all $\hat{A} \in \mathcal{O}_{S}$. For each $\hat{A} \in \mathcal{O}_{S}$, let $f_{\hat{A}}$ be the coordinate projection function $\Lambda \ri \sigma(\hat{A})$, that is, $f_{\hat{A}}(\lambda) \doteq \lambda_{\hat{A}}$ for all $\lambda \in \Lambda$. It follows from the definition of product $\sigma$-algebra that $f_{\hat{A}}$ is a measurable function \cite{tao2011measure}. For each $\Delta \subset \sigma(\hat{A})$, denote by $\kappa_{(\cdot)}[\hat{A} \in \Delta]$ the indicator function of the set $f_{\hat{A}}^{-1}(\Delta)$. It means that 
    \begin{equation}
        \kappa_{\lambda}[\hat{A} \in \Delta] = \begin{cases}
            1 \ \text{if} \ \lambda_{\hat{A}} \in \Delta\\
            0 \ \text{otherwise}.
        \end{cases}
    \end{equation}
    The definition of Lebesgue integral ensures that, for any measure $\mu$ on $\mathcal{A}$,
    \begin{equation}
        \mu(f_{\hat{A}}^{-1}(\Delta)) = \int_{\Lambda} \kappa_{\lambda}[\hat{A} \in \Delta] \ \mu(d\lambda).
    \end{equation}
    
    Thus far, we have constructed a space of ontic states $\boldsymbol{\Lambda} \equiv (\Lambda,\mathcal{A})$ and deterministic Markov kernels $\kappa_{(\cdot )}[\hat{A} \in \cdot \ ]$ (equivalently, measurable functions $f_{\hat{A}}:\Lambda \ri \R$) representing observables. To conclude, we need to define probability measures representing states and recover Eqs.~\eqref{eq:probabilityOntolgoicalDef} and \eqref{eq:conditionalProbabilityModel}. So let $\hat{\rho}$ be a state. For any finite subset $\{\hat{A}_{1},\dots,\hat{A}_{m}\}$ of $\mathcal{O}_{S}$, let $P_{\hat{\rho}}[(\hat{A}_{1},\dots,\hat{A}_{m}) \in \cdot \ ]$ be the joint distribution of  $\hat{A}_{1},\dots,\hat{A}_{m}$  in the state $\hat{\rho}$ (see Section~\ref{sec:compatibility}, more specifically Eqs.~\eqref{eq:finiteSequentialMeasure} and \eqref{eq:standardJoint}). Let $\hat{A}_{n_{1}},\dots,\hat{A}_{n_{M}}$ be a sub-sequence of $\hat{A}_{1},\dots,\hat{A}_{m}$. The joint distribution of $\hat{A}_{n_{1}},\dots,\hat{A}_{n_{M}}$ is the joint distribution of $\hat{A}_{1},\dots,\hat{A}_{m}$ marginalised over all observables in $\{\hat{A}_{1},\dots,\hat{A}_{m}\}$ except $\hat{A}_{n_{1}},\dots,\hat{A}_{n_{M}}$ --- this is known as the non-disturbance condition \cite{amaral2018graph}. Hence, it follows from the Kolmogorov extension theorem \cite{tao2011measure} that there exists a unique probability measure $\mu_{\hat{\rho}}$ on $\boldsymbol{\Lambda}$ such that, for any observables $\hat{A}_{1},\dots,\hat{A}_{m} \in \mathcal{O}_{S}$ and sets $\Delta_{i} \subset \sigma(\hat{A}_{i})$, $i=1,\dots,m$,
\begin{equation} \begin{split}
    P_{\hat{\rho}}[\hat{A}_{1} \in \Delta_{1}, \dots, \hat{A} \in \Delta_{m}] &=\label{eq:jointIntersection} \mu_{\hat{\rho}}(\bigcap_{i=1}^{m}f_{\hat{A}_{i}}^{-1}(\Delta_{i}))
    \\
    &= \int_{\Lambda} \prod_{i=1}^{m} \kappa_{\lambda}[\hat{A}_{i} \in \Delta_{i}] \ \mu_{\hat{\rho}}(d\lambda) 
\end{split}\end{equation}
(see Theorem 2.4.3 of Ref.~\cite{tao2011measure} for details). In particular, for any $\hat{A} \in \mathcal{O}_{S}$ and $\Delta \subset \sigma(\hat{A})$, Eq.~\eqref{eq:probabilityOntolgoicalDef} is satisfied. Finally, let's show that, for any state $\hat{\rho}$, any observable $\hat{A}$ and any set $\Delta \subset \sigma(\hat{A})$, the state $T_{[\hat{A} \in \Delta]}(\hat{\rho})$ is represented by the measure $\mu_{\hat{\rho}}$ conditioned on $f_{\hat{A}}^{-1}(\Delta)$, as in Eq.~\eqref{eq:conditionalProbabilityModel}. To begin with, it follows from Eqs.~\eqref{eq:finiteSequentialMeasure}, \eqref{eq:standardJoint} and \eqref{eq:jointIntersection} that, for any observables $\hat{A}_{1},\dots,\hat{A}_{m} \in \mathcal{O}_{S}$ and any  sets $\Delta_{i} \subset \sigma(\hat{A}_{i})$, $i=1,\dots,m$,
\begin{equation} \begin{split}
    P_{T_{[\hat{A} \in \Delta]}(\hat{\rho})}[\hat{A}_{1} \in \Delta_{1},\dots,\hat{A}_{m} \in \Delta_{m}] &= \frac{P_{\hat{\rho}}[\hat{A} \in \Delta,\hat{A}_{1} \in \Delta_{1},\dots,\hat{A}_{m} \in \Delta_{m}]}{P_{\hat{\rho}}[\hat{A} \in \Delta]}
    \\
    &=  \frac{\mu_{\hat{\rho}}(\bigcap_{i=1}^{m}f_{\hat{A}_{i}}^{-1}(\Delta_{i}) \cap f_{\hat{A}}^{-1}(\Delta))}{\mu_{\hat{\rho}}(f_{\hat{A}}^{-1}(\Delta))}
    \\
    &\equiv \mu_{\hat{\rho}}(\bigcap_{i=1}^{m}f_{\hat{A}_{i}}^{-1}(\Delta_{i}) \vert f_{\hat{A}}^{-1}(\Delta)),
\end{split}\end{equation}
where, as usual, $\mu_{\hat{\rho}}( \ \cdot \  \vert f_{\hat{A}}^{-1}(\Delta))$ denotes the measure $\mu_{\hat{\rho}}$ conditioned on $f_{\hat{A}}^{-1}(\Delta)$. On the other hand, Eq.~\eqref{eq:jointIntersection} ensures that 
\begin{equation}
    P_{T_{[\hat{A} \in \Delta]}(\hat{\rho})}[\hat{A}_{1} \in \Delta_{1},\dots,\hat{A}_{m} \in \Delta_{m}] = \mu_{T_{[\hat{A} \in \Delta]}(\hat{\rho})}(\bigcap_{i=1}^{m}f_{\hat{A}_{i}}^{-1}(\Delta_{i})),
\end{equation}
therefore
\begin{equation}
    \mu_{T_{[\hat{A} \in \Delta]}(\hat{\rho})}(\bigcap_{i=1}^{m}f_{\hat{A}_{i}}^{-1}(\Delta_{i})) = \mu_{\hat{\rho}}(\bigcap_{i=1}^{m}f_{\hat{A}_{i}}^{-1}(\Delta_{i}) \vert f_{\hat{A}}^{-1}(\Delta)).
\end{equation}
Hence, it follows from the uniqueness of the measure given by the Kolmogorov extension theorem \cite{tao2011measure} (applied to the state $T_{[\hat{A} \in \Delta]}(\hat{\rho})$) that
\begin{equation}
    \mu_{T_{[\hat{A} \in \Delta]}(\hat{\rho})}( \ \cdot \ ) = \mu_{\hat{\rho}}( \ \cdot \ \vert f_{\hat{A}}^{-1}(\Delta)).
\end{equation}
It proves that $\tau_{[\hat{A} \in \Delta]}$ is necessarily be given by Eq.~\eqref{eq:conditionalProbabilityModel}, completing the proof.
\hfill\ $\square \medskip$

\noindent \textbf{Proof of Lemma~\ref{lemma:variationDistance}:} 
Let $\mathfrak{M} \equiv (\boldsymbol{\Lambda},\Theta,\Psi)$ be a deterministic (state updating) ontological model for $\mathcal{O}_{S}$. Let $\mathcal{S}_{\Lambda}$ be the set of all probability measures on $\boldsymbol{\Lambda}$, and let $\mu \in \mathcal{S}_{\Lambda}$. For any $\hat{A} \in \mathcal{O}_{S}$  and $\Delta \subset \sigma(\hat{A})$, set $\Omega_{\Delta} \doteq f_{\hat{A}}^{-1}(\Delta)$ and $\Omega_{\Delta}^{C} \doteq \Lambda \backslash \Omega_{\Delta}$.

Let $\nu$ be a probability measure whose support lies in $\Omega_{\Delta}$, i.e., $1=P_{\nu}[\hat{A} \in \Delta] \equiv  \nu(\Omega_{\Delta})$. Define $\eta:\mathcal{A} \ri [-1,1]$ by
\begin{equation}
    \eta(\Omega) \doteq (\nu - \mu)(\Omega)
\end{equation}
for each $\Omega \in \Omega$. This is what is called a signed measure in the literature \cite{folland1999real,cohn2013measure}. Note that $\eta(\Omega) = -\mu(\Omega) \leq 0$ whenever $\Omega \subset \Omega_{\Delta}^{C}$. 

The well-known Hahn decomposition theorem \cite{folland1999real} ensures the existence of a partition $\Omega_{-}',\Omega_{+}' \in \mathcal{A}$ of $\Lambda$ such that $\eta(\Omega) \leq 0$ if $\Omega \subset \Omega_{-}'$ and  $\eta(\Omega) \geq 0$ whenever $\Omega \subset \Omega_{+}'$. Since $\Omega_{+}' \cap \Omega_{\Delta}^{C} \subset \Omega_{\Delta}^{C}$, we have $\eta(\Omega_{+}' \cap \Omega_{\Delta}^{C}) \leq 0$. On the other hand, $\Omega_{+}' \cap \Omega_{\Delta}^{C} \subset \Omega_{+}$ implies that   $\eta(\Omega_{+}' \cap \Omega_{\Delta}^{C}) \geq 0$. Therefore, $\eta(\Omega_{+} \cap \Omega_{\Delta}^{C}) = 0$. Define $\Omega_{+} \doteq \Omega_{+}' \cap \Omega_{\Delta}$ and $\Omega_{-} \doteq \Lambda \backslash \Omega_{+} = \Omega_{-}' \cup \Omega_{\Delta}^{C}$. Then $\Omega_{+},\Omega_{-} \in \mathcal{A}$ is a partition of $\Lambda$ such that $\eta(\Omega) \geq 0$ whenever $\Omega \subset \Omega_{+}$ and $\eta(\Omega) \leq 0$ if $\Omega \subset \Omega_{-}$. In fact,  $\Omega \subset \Omega_{+}$ implies $\Omega \subset \Omega_{+}'$, thus $\eta(\Omega) \geq 0$. On the other hand, if $\Omega  \subset \Omega_{-}'$ then $\eta(\Omega) = \eta(\Omega \cap \Omega_{\Delta}^{C}) + \eta(\Omega \cap \Omega_{\Delta})$, in which $\Omega \cap \Omega_{\Delta}^{C} \subset \Omega_{\Delta}^{C}$ and $\Omega \cap \Omega_{\Delta} \subset \Omega_{-}^{'}\backslash \Omega_{\Delta}^{C} \subset \Omega_{-}'$, thus $\eta(\Omega) \leq 0$. Note that, by construction, we have $\Omega_{+} \subset \Omega_{\Delta}$.

Define $\eta^{+}(\Omega) \doteq \eta (\Omega \cap \Omega_{+})$ and $\eta^{-}(\Omega) \doteq -\eta(\Omega \cap \Omega_{-})$ for each $\Omega \in \mathcal{A}$. More explicitly,
\begin{equation}
    \begin{split}
        \eta^{+}(\Omega) \doteq (\nu - \mu)(\Omega \cap \Omega_{+}),\\
        \eta^{-}(\Omega) \doteq (\mu - \nu)(\Omega \cap \Omega_{-}).
    \end{split}
\end{equation}
$\eta^{+}$ and $\eta^{-}$ are measures on $\boldsymbol{\Lambda}$ \cite{folland1999real}. We have $\eta = \eta^{+} - \eta^{-}$, which is known as the Jordan decomposition of $\eta$ \cite{folland1999real}. It is well-known that the total variation distance between $\nu$ and $\mu$ (Eq.~\eqref{eq:variationDistance}) equals half the total variation of $\eta \doteq \nu - \mu$, which in turn is the number $(\eta^{+} + \eta^{-})(\Lambda)$ \cite{klenke2014probability}. Therefore,
\begin{equation}
    \begin{split}
        2\Vert \nu - \mu \Vert &=  (\eta^{+} + \eta^{-})(\Lambda) = \eta^{+}(\Omega_{+}) + \eta^{-}(\Omega_{-})\\
        &= [(\nu-\mu)(\Omega_{\Delta}) - (\nu-\mu)(\Omega_{\Delta}\backslash\Omega_{+})] + \eta^{-}(\Omega_{-})\\
        &\geq (\nu-\mu)(\Omega_{\Delta})  + \eta^{-}(\Omega_{-}) = 1-\mu(\Omega_{\Delta}) + (\mu - \nu)(\Omega_{-})\\
        &=  1-\mu(\Omega_{\Delta}) + \mu(\Lambda \backslash \Omega_{\Delta}) + \mu(\Omega_{-} \cap \Omega_{\Delta}) - \nu(\Omega_{-} \cap \Omega_{\Delta})\\
        &=2 - 2\mu(\Omega_{\Delta})  + \eta^{-}(\Omega_{\Delta})\\
        &\geq\label{eq:distanceFromSupported} 2 - 2\mu(\Omega_{\Delta}).
    \end{split}
\end{equation}

Next, consider $\mu$ conditioned on $\Omega_{\Delta}$, namely the measure
\begin{equation}
    \tau_{[\hat{A} \in \Delta]}(\mu) \equiv \mu(\  \cdot \ |\Omega_{\Delta}) = \frac{\mu( \ \cdot \ \cap \Omega_{\Delta})}{\mu(\Omega_{\Delta})}.
\end{equation}
Clearly, $\tau_{[\hat{A} \in \Delta]}(\mu)$ assigns probability $1$ to $\Omega_{\Delta}$. For each $\Omega \subset \Omega_{\Delta}$ we have $\mu(\Omega \vert \Omega_{\Delta}) \geq \mu(\Omega)$, whereas $0 =\mu(\Omega\vert \Omega_{\Delta})\leq \mu(\Omega)$ whenever $\Omega \subset \Lambda \backslash \Omega_{\Delta}$. Hence, $\Omega_{\Delta},\Omega_{\Delta}^{C}$ is a Hahn decomposition of $\Lambda$ with respect to the signed measure $\tau_{[\hat{A} \in \Delta]}(\mu) - \mu$ \cite{folland1999real}. It leads us to
\begin{equation}
    \begin{split}
        2 \Vert \tau_{[\hat{A} \in \Delta]}(\mu) - \mu \Vert &= (\tau_{[\hat{A} \in \Delta]}(\mu) - \mu)(\Omega_{\Delta}) +  (\mu - \tau_{[\hat{A} \in \Delta]}(\mu))(\Omega_{\Delta}^{C})\\
        &= \mu(\Omega_{\Delta}|\Omega_{\Delta}) - \mu(\Omega_{\Delta}) + 1- \mu(\Omega_{\Delta}) - \mu(\Omega_{\Delta}^{C}|\Omega_{\Delta})
        \\
        &=\label{eq:distanceFromConditioned} 2 - 2\mu(\Omega_{\Delta}).
    \end{split}
\end{equation}

Finally, let $\mathcal{S}_{\Lambda}(\hat{A} \in \Delta)$ be the set of all probability measures on $\boldsymbol{\Lambda}$ such that  $P_{\nu}[\hat{A} \in \Delta] = 1$. It follows from Eqs.~\eqref{eq:distanceFromSupported} and \eqref{eq:distanceFromConditioned} that, for each $\eta \in \mathcal{S}_{\Lambda}(\hat{A} \in \Delta)$,
\begin{equation}
     \Vert \tau_{[\hat{A} \in \Delta]}(\mu) - \mu \Vert  \leq \Vert \nu - \mu \Vert. 
\end{equation}
We have $\tau_{[\hat{A} \in \Delta]}(\mu) \in \mathcal{S}_{\Lambda}(\hat{A} \in \Delta)$, thus
    \begin{equation}
    \begin{split}
        \Vert \tau_{[\hat{A} \in \Delta]}(\mu) - \mu\Vert &= \min\{\Vert \nu - \mu\Vert: \nu \in \mathcal{S}_{\Lambda}(\hat{A} \in \Delta)\}\\
        &= \min\{\Vert \nu - \mu\Vert: \nu \in \mathcal{S}_{\Lambda}, P_{\nu}[\hat{A} \in \Delta] = 1\}.
    \end{split}
    \end{equation}
This completes the proof.
\hfill\ $\square \medskip$

\end{document}